\documentclass[twoside,11pt]{article}

\usepackage{amsmath,amsfonts,amssymb,amsthm}
\usepackage{MyStyleFile}
\usepackage{subcaption,natbib,authblk,setspace,xcolor,graphicx,url,enumerate}
\graphicspath{{Plots1/}}

\usepackage[left=2.5cm,right=2.5cm,top=3cm,bottom=2.5cm]{geometry}

\usepackage{hyperref}
\hypersetup{
     colorlinks=true,
     linkcolor=black,
     citecolor=blue,
     filecolor=blue,
     urlcolor=blue,
}

\begin{document}

\title{\Large\sc On High-dimensional Modifications\\ of Some Graph-based Two-sample Tests}

\date{}

\author[a]{Soham Sarkar \thanks{sohamsarkar1991@gmail.com}}
\author[b]{Rahul Biswas \thanks{rbiswas1@uw.edu}}
\author[a]{Anil K. Ghosh \thanks{akghosh@isical.ac.in}}

\affil[a]{Theoretical Statistics and Mathematics Unit, Indian Statistical Institute \protect\\ 203, B. T. Road, Kolkata 700108, India.}
\affil[b]{Department of Statistics, University of Washington \protect\\ Seattle, Washington 98195, U.S.A.}

\maketitle

\vspace{-0.3in}
\begin{abstract}%
Testing for the equality of two high-dimensional distributions is a challenging problem, and this becomes even more challenging when the sample size is small. Over the last few decades, several graph-based two-sample tests have been proposed in the literature, which can be used for data of arbitrary dimensions. Most of these test statistics are computed using pairwise Euclidean distances among the observations. But, due to concentration of pairwise Euclidean distances, these tests have poor performance in many high-dimensional problems. Some of them can have powers even below the nominal level when the scale-difference between two distributions dominates the location-difference. To overcome these limitations, we introduce a new class of dissimilarity indices and use it to modify some popular graph-based tests. These modified tests use the distance concentration phenomenon to their advantage, and as a result, they outperform the corresponding tests based on the Euclidean distance in a wide variety of examples. We establish the high-dimensional consistency of these modified tests under fairly general conditions. Analyzing several simulated as well as real data sets, we demonstrate their usefulness in high dimension, low sample size situations.
\vspace{0.05in}

\end{abstract}

\onehalfspacing
\section{Introduction}
\label{sec:intro}

Let ${\cal X}_m=\{\xvec_{1},\ldots,\xvec_{m}\}$ and ${\cal Y}_n=\{\yvec_{1},\ldots,\yvec_{n}\}$ be two sets of independent observations from $d$-dimensional continuous distributions $F$ and $G$, respectively. In the two-sample problem, we use these observations to test the null hypothesis ${\cal H}_0:F=G$ against the alternative hypothesis ${\cal H}_A:F \neq G$. This problem is well-investigated, and several tests are available for it. Interestingly, many of these tests are based on pairwise Euclidean distances among the observations. Under some mild conditions, \cite{MPB96} showed that for $\Xvec_1,\Xvec_2 \stackrel{i.i.d.}{\sim}F$ and $\Yvec_1,\Yvec_2 \stackrel{i.i.d.}{\sim}G$, $\|\Xvec_1-\Xvec_2\|$, $\|\Yvec_1-\Yvec_2\|$ and $\|\Xvec_1-\Yvec_1\|$ have the same distribution if and only if $F$ and $G$ are identical. So, pairwise Euclidean distances contain useful information about the difference between two distributions, and these distances can be easily computed even in high dimensions.  Because of these reasons, pairwise Euclidean distances have been extensively used for the construction of two-sample tests, which are applicable to high dimension, low sample size (HDLSS) data.

The existing tests based on pairwise Euclidean distances can be broadly categorized into two groups: (i) tests based on averages of three types ($\Xvec\Xvec$, $\Xvec\Yvec$ and $\Yvec\Yvec$) of pairwise distances and (ii) tests based on graphs. \cite{BF04} and \cite{SR04} were the first to construct tests based on averages of pairwise Euclidean distances. \cite{AZ05} considered tests based on averages of functions of pairwise distances, which they called the energy distance. The test based on maximum mean discrepancy statistic by \cite{Gretton12} can be viewed as a kernelized version of the test proposed by \cite{BF04}. Other two-sample tests based on averages of pairwise Euclidean distances include \cite{BF10,SR13,BG14} and \cite{T17}.

The graph-based tests consider an edge-weighted complete graph ${\cal G}$ on the vertex set ${\cal Z}_N={\cal X}_m\cup{\cal Y}_n$ (here $N=m+n$ is the total sample size), where the Euclidean distance between two vertices is taken to be the weight associated with the edge connecting them. Different tests consider different sub-graphs of ${\cal G}$ and look at their topologies. The deviation of the topology of a sub-graph from the one expected under
${\cal H}_0$ is used to construct the test statistic. \cite{FR79} were the first to develop such graph-based tests for multivariate data. They proposed multivariate generalizations of the Kolmogorov-Smirnov test and the  Wald-Wolfowitz run test using the minimum spanning tree (MST) of ${\cal G}$. \cite{BMG14} used the shortest Hamiltonian path (SHP) on ${\cal G}$, instead of MST, to construct another multivariate run test. \cite{R05} constructed the cross-match test using $\lfloor N/2 \rfloor$ disconnected edges of ${\cal G}$ (here $\lfloor t \rfloor$ denotes the largest integer smaller than or equal to $t$) for which the total edge weight is minimum. \cite{LM11} considered all cliques of size $3$ to construct their test statistic. Recently, \cite{CF17} also constructed some tests using graph-theoretic ideas. The tests based on nearest-neighbor type coincidences \citep[see, e.g.,][]{S86,H88,HT02,MBG15} can be viewed as tests based on directed sub-graphs of ${\cal G}$ (see the discussion on NN test in Page 3).

Recently, \cite{SG18} showed that due to concentration of pairwise Euclidean distances, the tests based on averages of pairwise distances can have very low powers in many high-dimensional examples. Instead of the Euclidean distance, they suggested to use distance functions of the form $\varphi_{h,\psi}(\uvec,\vvec)= h\{\frac{1}{d}\sum_{q=1}^{d} \psi(|u^{(q)}-v^{(q)}|)\}$ for suitably chosen strictly increasing functions $h,\psi:[0,\infty)\rightarrow[0, \infty)$ with $h(0)=\psi(0)=0$. For their choices of $h$ and $\psi$, tests based on averages of pairwise $\varphi_{h,\psi}$-distances outperformed those based on pairwise Euclidean distances in many examples \citep[see][]{SG18}. Naturally, one would like to know whether the graph-based tests constructed using pairwise Euclidean distances also have similar problems in high dimensions. For this investigation, we consider two simple examples.

\begin{example}
\label{example1}
$F$ and $G$ are Gaussian with the same mean and diagonal dispersion matrices $\Lambdat_{1,d}$ and $ \Lambdat_{2,d}$, respectively. The first $d/2$ diagonal elements of $\Lambdat_{1,d}$ are $1$ and the rest are $2$, whereas for $\Lambdat_{2,d}$, the first $d/2$ diagonal elements are $2$ and the rest are $1$.
\end{example}
\begin{example}
\label{example2}
For $\Xvec=(X^{(1)},\ldots,X^{(d)})^{\top} \sim F$, $X^{(1)},\ldots,X^{(d)}$ are i.i.d. as ${\cal N}(0,5)$, while for $\Yvec=(Y^{(1)},\ldots,Y^{(d)})^{\top} \sim G$, $Y^{(1)},\ldots,Y^{(d)}$ are i.i.d. $t_{5}(0,3)$. Here ${\cal N}(\mu,\sigma^2)$ denotes the normal distribution with mean $\mu$ and variance $\sigma^2$, and $t_{\nu}(\mu,\sigma^2)$ denotes the Student's $t$-distribution with $\nu$ degrees of freedom, location $\mu$ and scale $\sigma$.
\end{example}
\vspace{-0.05in}

For both of these examples, we performed our experiments with $d=2^{i}$ for $i=1,\ldots,10$. For different values of $d$, we generated $20$ observations from each distribution and used them to test ${\cal H}_0:F=G$ against ${\cal H}_A:F \neq G$. We repeated each experiment $500$ times and estimated the power of a test by the proportion of times it rejected ${\cal H}_0$. Figures~\ref{fig:Ex1} and \ref{fig:Ex2} show the observed powers for four popular graph-based tests (of 5\% nominal level), namely, the test based on  nearest-neighbors \citep{S86,H88}, the multivariate run test based on MST \citep{FR79}, the multivariate run test based on SHP \citep{BMG14} and the cross-match test based on optimal non-bipartite matching \citep{R05}. Henceforth, they will be referred to as the NN test, the MST-run test, the SHP-run test and the NBP test, respectively. Brief descriptions of these four tests are given below.

{\bf NN test} \citep{S86,H88}: Consider the edge-weighted complete graph ${\cal G}$ on vertex set ${\cal Z}_N$, where the edge-weights are defined using pairwise Euclidean distances. Assume that an undirected edge $(\uvec,\vvec)$ in ${\cal G}$ corresponds to two directed edges $(\overrightarrow{\uvec,\vvec})$ and $(\overrightarrow{\vvec, \uvec})$. Now, for a fixed $k < N$, consider the sub-graph ${\cal T}_{k}$, which contains an edge $(\overrightarrow{\uvec,\vvec})$ if and only if  $\vvec$ is among the first $k$ nearest-neighbors (in terms of the Euclidean distance) of $\uvec$. Clearly, ${\cal T}_k$ contains $Nk$ directed edges. The NN test uses the test statistic $T_{NN} = \frac{1}{Nk} \sum_{(\overrightarrow{\uvec,\vvec}) \in {\cal T}_k} \bI(\uvec,\vvec)$, where $\bI(\uvec,\vvec)$ is an indicator variable that takes the value $1$ if $\uvec$ and $\vvec$ are from the same distribution. It rejects ${\cal H}_0$ for large values of $T_{NN}$. A more familiar expression of this test statistic is $T_{NN} = \frac{1}{Nk} \big[\sum_{i=1}^m \sum_{r=1}^k \bI_r(\xvec_i) +\sum_{i=1}^n \sum_{r=1}^k \bI_r(\yvec_i)\big]$, where $\bI_r(\zvec)$ is an indicator variable that takes the value $1$ if $\zvec$ and its $r$-th nearest-neighbor (in terms of the Euclidean distance) come from the same distribution.

{\bf MST-run test} \citep{FR79}: Unlike the NN test, this test is based on an undirected sub-graph of ${\cal G}$. Let {${\cal M}$} {be the minimum spanning tree (MST)} of {${\cal G}$}. The MST-run test uses the test statistic $T_{MST} = 1 + \sum_{i=1}^{N-1} \lambda^{\cal M}_i$, where $\lambda^{\cal M}_i$ is an indicator variable that takes the value $1$ if and only if the $i$-th edge $(i=1,\ldots,N-1)$ of ${\cal M}$ connects two observations from different distributions. The null hypothesis ${\cal H}_0$ is rejected for small values of $T_{MST}$.

{\bf SHP-run test} \citep{BMG14}: Instead of MST, this test uses the shortest Hamiltonian path (SHP). Let {${\cal S}$} {be the SHP on ${\cal G}$. The number of runs along ${\cal S}$ is computed as $T_{SHP} = 1 + \sum_{i=1}^{N-1} \lambda^{\cal S}_i$, where the indicator $\lambda^{\cal S}_i$ takes the value $1$ if and only if the $i$-th edge of ${\cal S}$ connects two observations from different distributions. The SHP-run test rejects ${\cal H}_0$ for small values of $T_{SHP}$.

{\bf NBP test} \citep{R05}: {It uses the optimal non-bipartite matching algorithm \citep[see, e.g.,][]{Lu11} to find $\lfloor N/2\rfloor$ disconnected edges (i.e., no two edges share a common vertex) in ${\cal G}$ such that the total weight of the edges is minimum. Let ${\cal C}=\{(\uvec_{i},\vvec_{i}): ~i=1,\ldots,\lfloor N/2 \rfloor\}$ be the collection of these edges. The NBP test rejects ${\cal H}_0$ for small values of the test statistic $T_{NBP} = \sum_{i=1}^{N/2} \lambda^{\cal C}_i$, where $\lambda^{\cal C}_i$ is an indicator variable that takes the value $1$ if and only if $\uvec_{i}$ and $\vvec_{i}$ are from two different distributions.

\begin{figure}[h]
\begin{subfigure}{0.45\textwidth}
\centering
\subcaption{NN tests}
\includegraphics[height=2.20in,width=\linewidth]{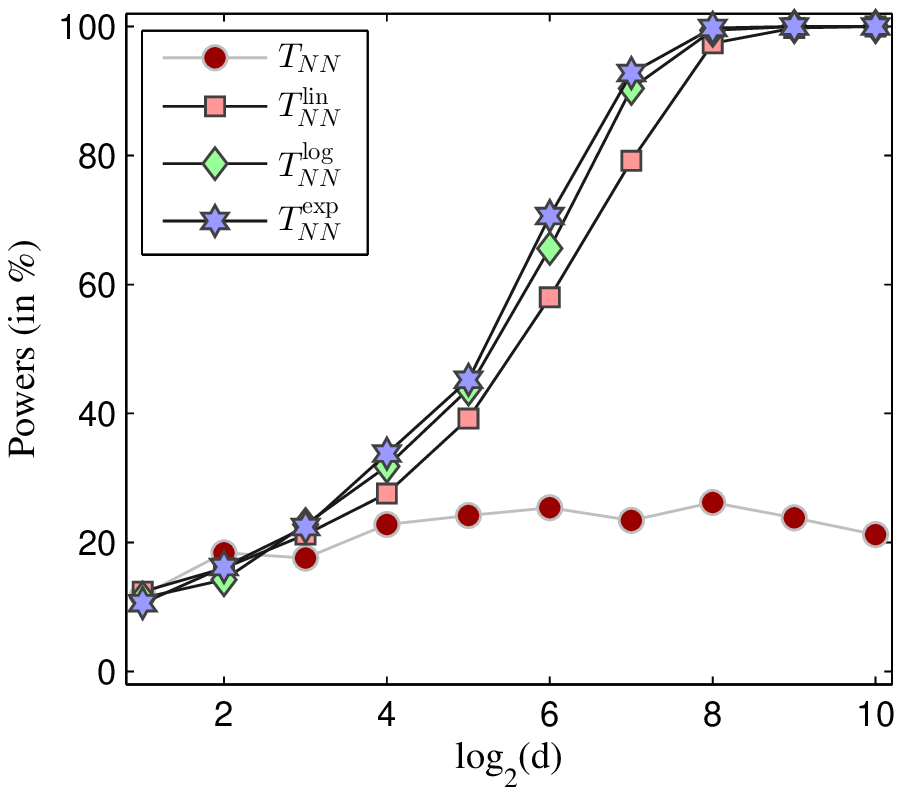}
\end{subfigure}%
\hspace{0.09\textwidth}%
\begin{subfigure}{0.45\textwidth}
\centering
\subcaption{MST-run tests}
\includegraphics[height=2.20in,width=\linewidth]{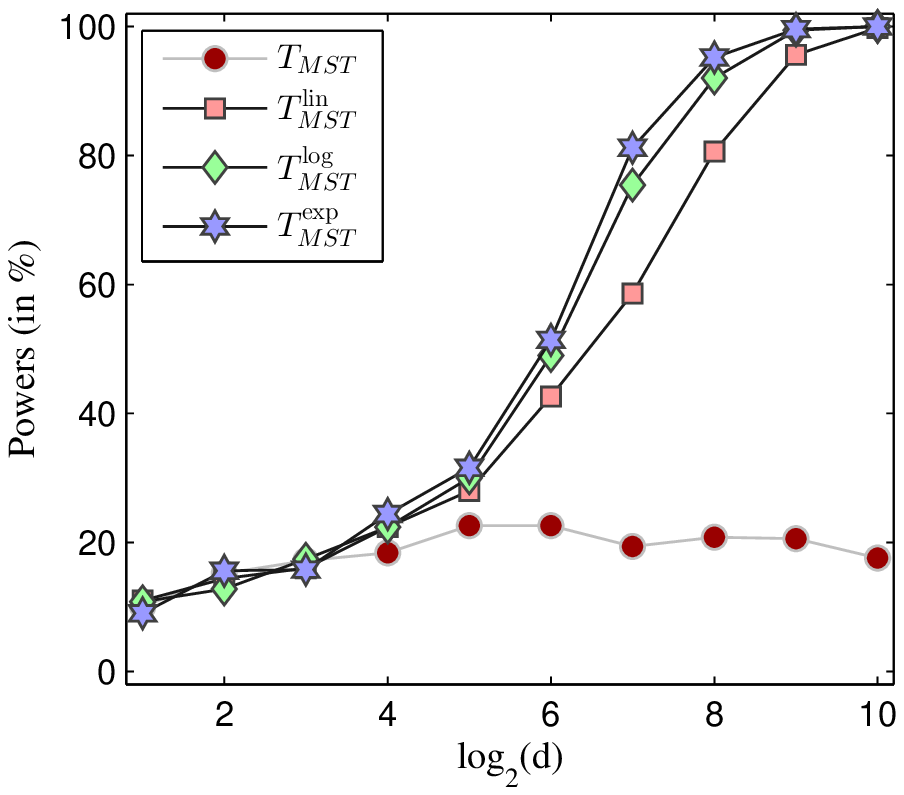}
\end{subfigure}
\begin{subfigure}{0.45\textwidth}
\centering
\subcaption{SHP-run tests}
\includegraphics[height=2.20in,width=\linewidth]{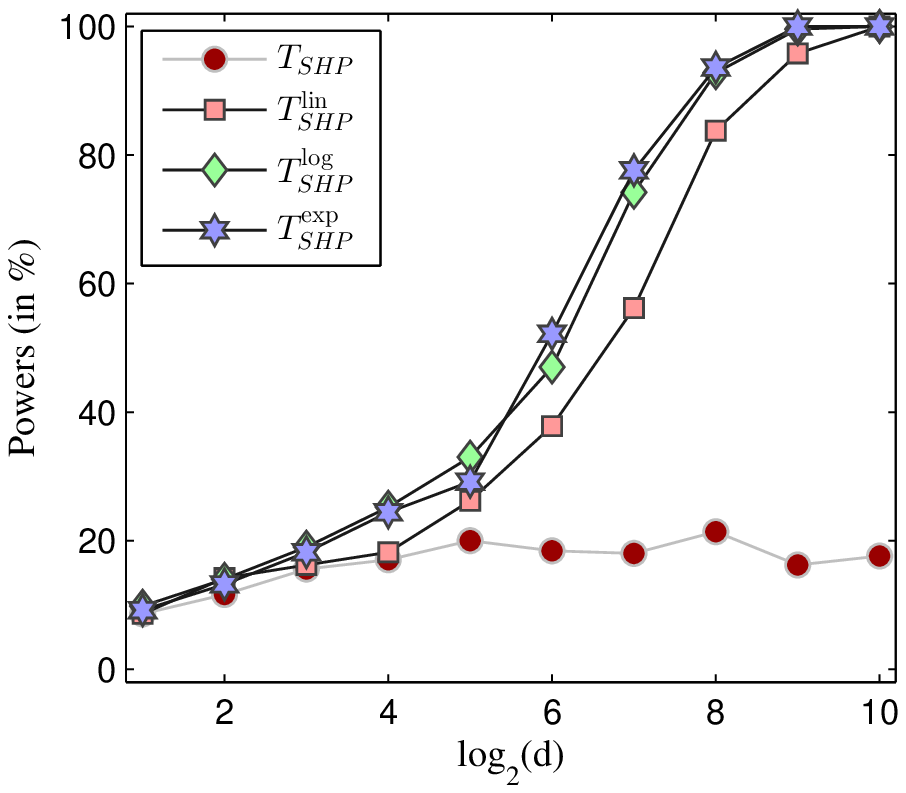}
\end{subfigure}%
\hspace{0.09\textwidth}%
\begin{subfigure}{0.45\textwidth}
\centering
\subcaption{NBP tests}
\includegraphics[height=2.20in,width=\linewidth]{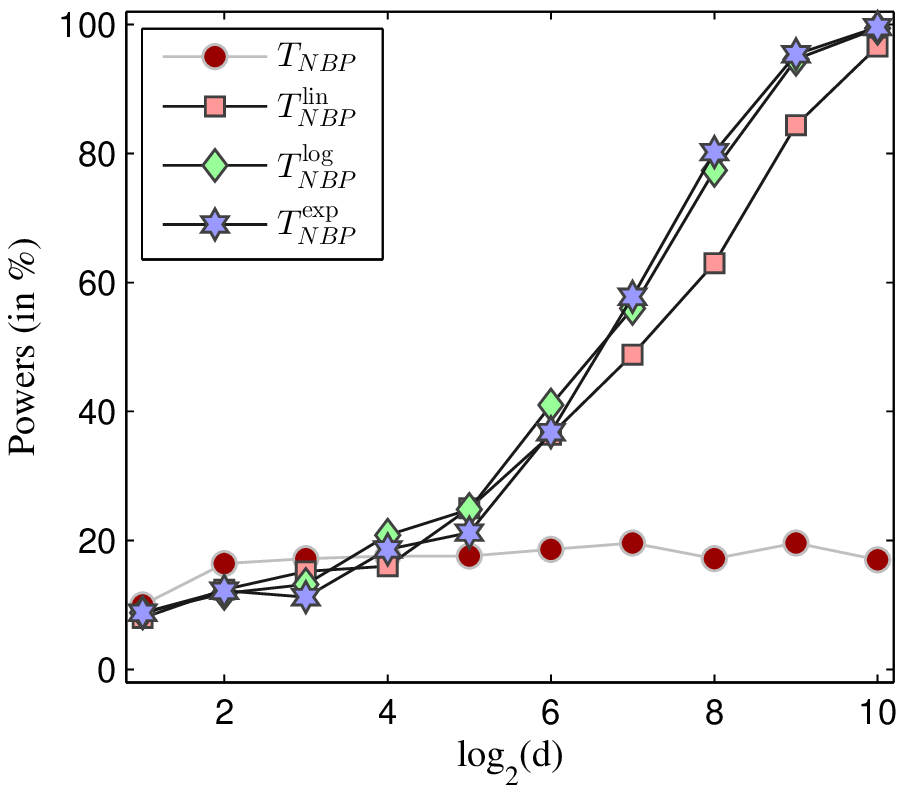}
\end{subfigure}
\caption{Powers of NN, MST-run, SHP-run and NBP tests in Example \ref{example1}.}
\label{fig:Ex1}
\end{figure}

The SHP-run test and the NBP test are distribution-free. For the NN test and the MST-run test, throughout this article, we use conditional tests based on 1000 random permutations. For the NN test, we use $k=3$ for all numerical work since it has been reported to perform well in the literature \citep[see, e.g.,][]{S86}.

\begin{figure}[h]
\begin{subfigure}{0.45\textwidth}
\centering
\subcaption{NN tests}
\includegraphics[height=2.20in,width=\linewidth]{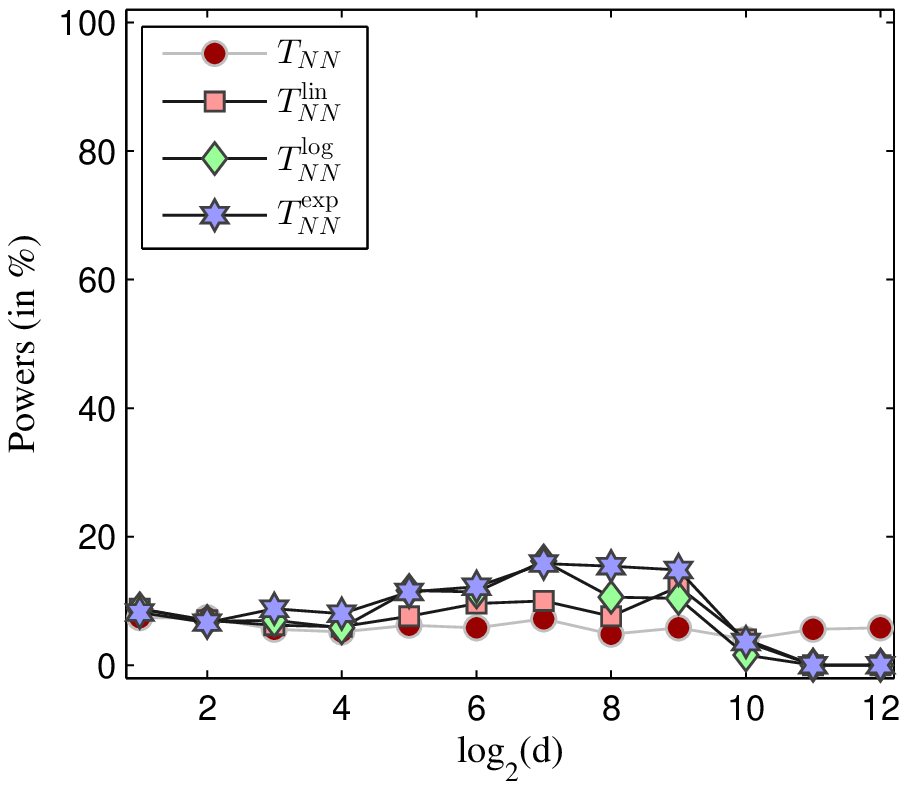}
\end{subfigure}%
\hspace{0.09\textwidth}%
\begin{subfigure}{0.45\textwidth}
\centering
\subcaption{MST-run tests}
\includegraphics[height=2.20in,width=\linewidth]{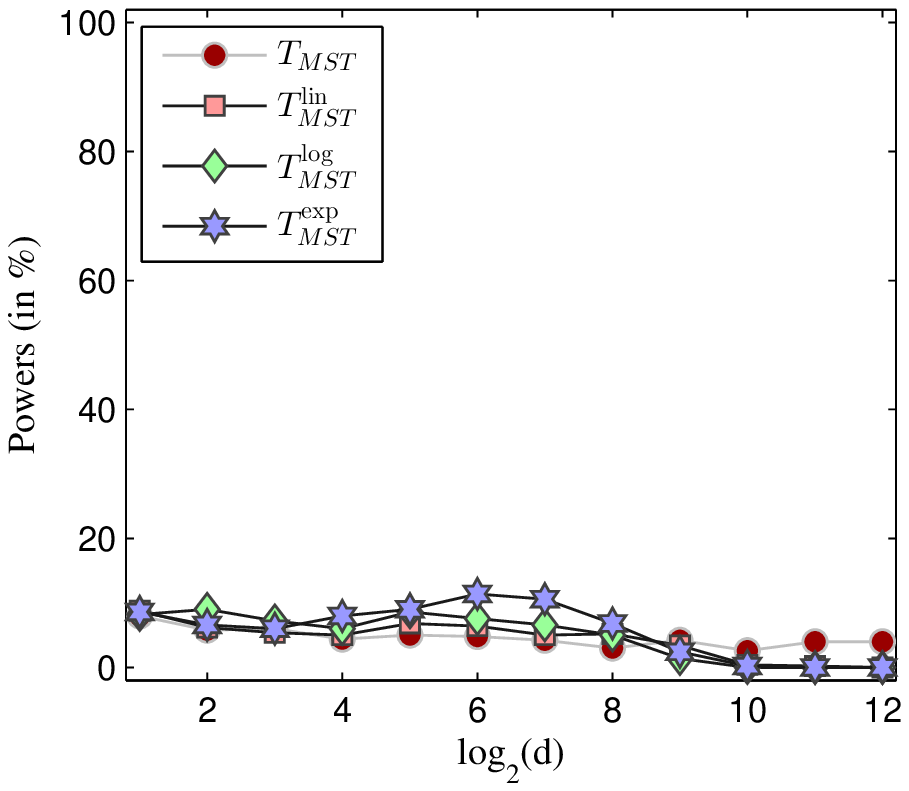}
\end{subfigure}
\begin{subfigure}{0.45\textwidth}
\centering
\subcaption{SHP-run tests}
\includegraphics[height=2.20in,width=\linewidth]{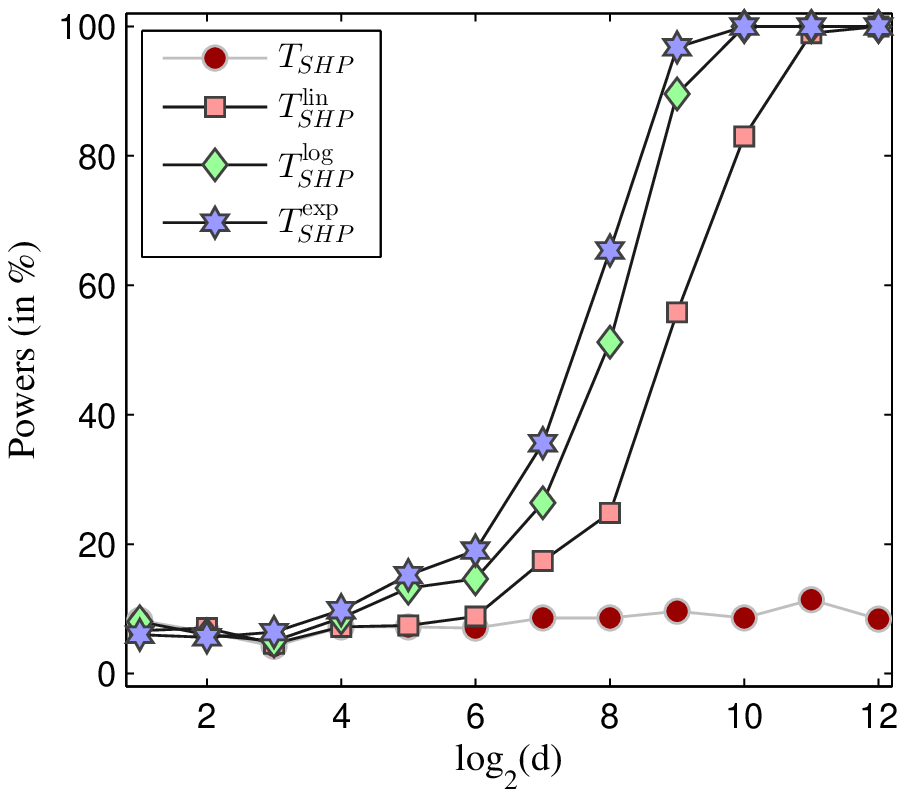}
\end{subfigure}%
\hspace{0.09\textwidth}%
\begin{subfigure}{0.45\textwidth}
\centering
\subcaption{NBP tests}
\includegraphics[height=2.20in,width=\linewidth]{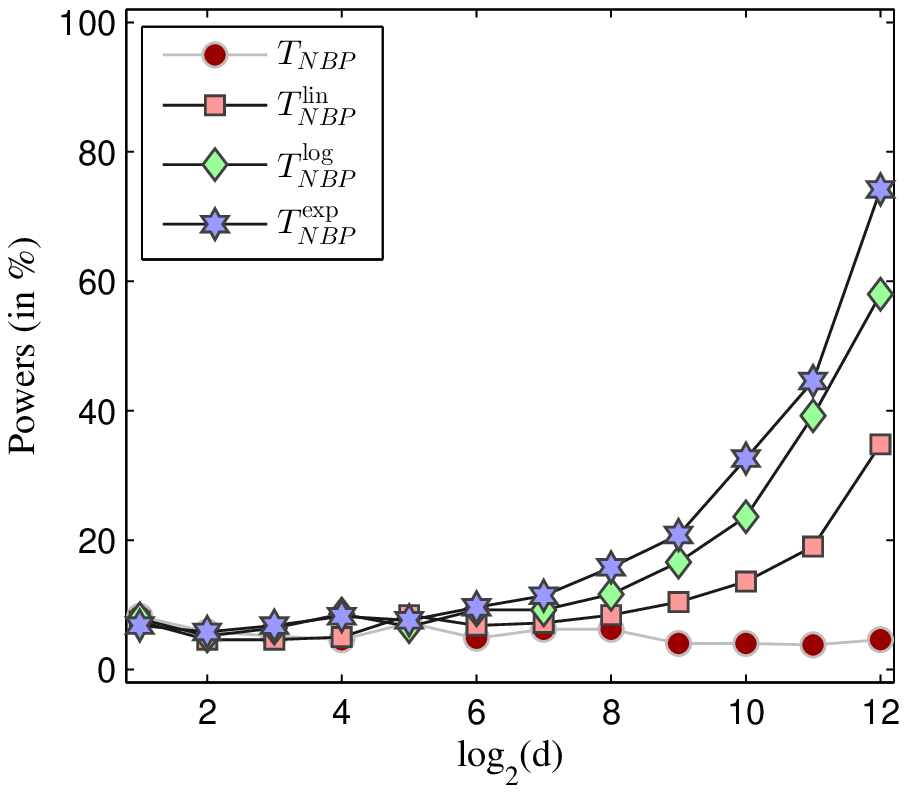}
\end{subfigure}
\caption{Powers of NN, MST-run, SHP-run and NBP tests in Example~\ref{example2}.}
\label{fig:Ex2}
\end{figure}

Figures~\ref{fig:Ex1} and \ref{fig:Ex2} clearly show that all these tests based on pairwise Euclidean distances had poor performance in Examples~\ref{example1} and \ref{example2}. Note that in both of these examples, each measurement variable has different distributions under $F$ and $G$. So, each of them carries signal against ${\cal H}_0$. Therefore, the power of a test is expected to increase to unity as the dimension increases. But we did not observe that for these tests based on the Euclidean distance. Now, one may be curious to know what happens to these tests if the Euclidean distance is replaced by the distance function $\varphi_{h, \psi}$ (i.e., the edge-weights in ${\cal G}$ are defined using $\varphi_{h,\psi}$) as in \cite{SG18}. Here we consider three choices of $\psi$, namely, $\psi_1(t)=t$, $\psi_2(t)=\log(1+t)$ and $\psi_3(t)=1-\exp(-t)$, with $h(t)=t$ in all three cases. Note that these choices satisfy the desirable properties mentioned in \cite{SG18}. The curves corresponding to ${T}^{\rm lin}$, ${T}^{\log}$ and ${T}^{\exp}$ in Figures~\ref{fig:Ex1} and \ref{fig:Ex2} represent the powers of the tests based on $\varphi_{\psi,h}$ with $\psi_1$, $\psi_2$ and $\psi_3$, respectively. These tests had excellent performance in Example~\ref{example1}. Their powers converged to unity as the dimension increased. Modified SHP-run tests based on $\varphi_{h,\psi}$ had similar behavior in Example~\ref{example2} as well. In this example, powers of modified NBP tests also increased with the dimension, but those of modified NN and MST-run tests dropped down to zero as the dimension increased.

In the next section, we investigate the reasons behind the contrasting behavior of these tests in Examples~\ref{example1} and \ref{example2}. In order to overcome the limitations of NN and MST-runs tests, in Section~\ref{sec:MADD}, we construct a new class of dissimilarity indices and modify NN and MST-run tests using them. High-dimensional behavior of the resulting tests are also studied under appropriate regularity conditions. Some simulated and real data sets are analyzed in Section~\ref{sec:simulation} to study the empirical performance of the tests. Section~\ref{sec:remarks} contains a brief summary of the work and ends with a discussion on possible directions for future research. All proofs and mathematical details are given in the Appendix.

\section{High-dimensional behavior of the tests based on the Euclidean distance and $\varphi_{h,\psi}$}
\label{sec:varphi}

To properly understand the high-dimensional behavior of different {graph-based} tests used in Section~\ref{sec:intro}, we consider another example (call it {\bf Example 3}) involving two multivariate normal distributions ${\cal N}_d({\bf 0}_d,{\bf I}_d)$ and ${\cal N}_d(0.2 {\bf 1}_d, \gamma^{-1} {\bf I}_d)$. Here ${\cal N}_d(\muvec,\sigmat)$ denotes the $d$-variate normal distribution with mean $\muvec$ and dispersion matrix $\sigmat$, ${\bf 0}_d=(0,0,\ldots,0)^{\top} \in {\mathbb {R}}^d$, ${\bf 1}_d=(1,1,\ldots,1)^{\top} \in {\mathbb {R}}^d$, and ${\bf I}_d$ denotes the $d \times d$ identity matrix. Keeping $d$ fixed at $250$, we generated $20$ observations from each distribution and repeated the experiment $500$ times to estimate the powers of different tests, which are shown in Figure~\ref{fig:ExA} as functions of $\gamma$. In this example, as $\gamma$ increases, the separation between the two distributions also increases. So, the power of any reasonable test is expected to increase with $\gamma$. We observed this for all versions of SHP-run and NBP tests, but that was not the case for NN and MST-run tests. {In fact}, their powers dropped down to zero as $\gamma$ increased.

\begin{figure}[h]
\begin{subfigure}{0.45\textwidth}
\centering
\subcaption{NN tests}
\includegraphics[height=2.20in,width=\linewidth]{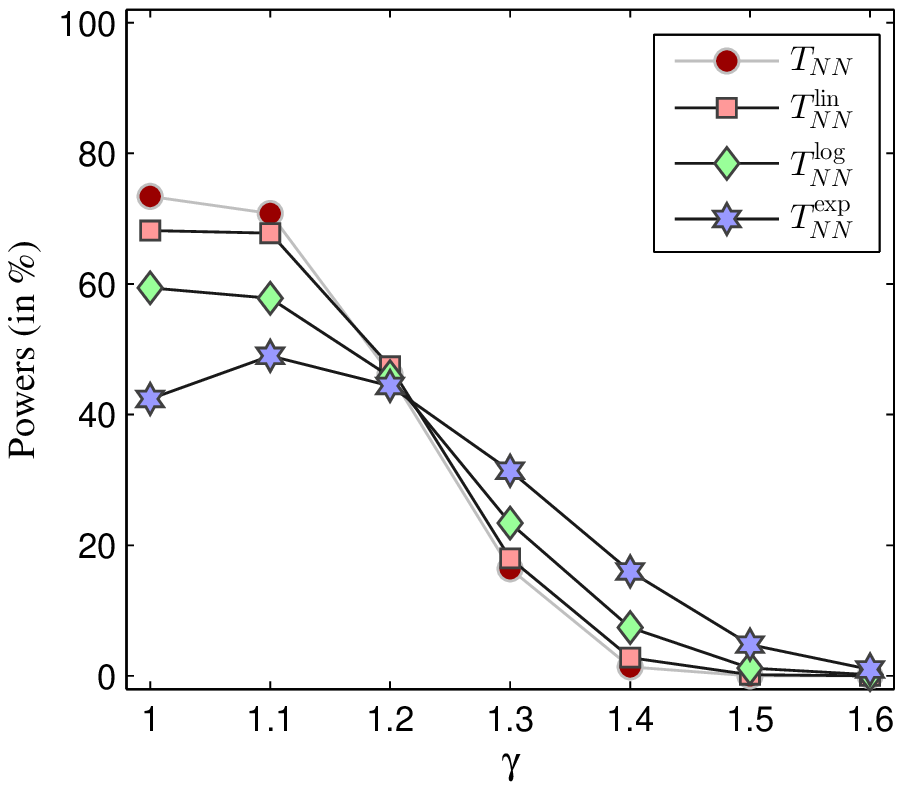}
\end{subfigure}%
\hspace{0.09\textwidth}%
\begin{subfigure}{0.45\textwidth}
\centering
\subcaption{MST-run tests}
\includegraphics[height=2.20in,width=\linewidth]{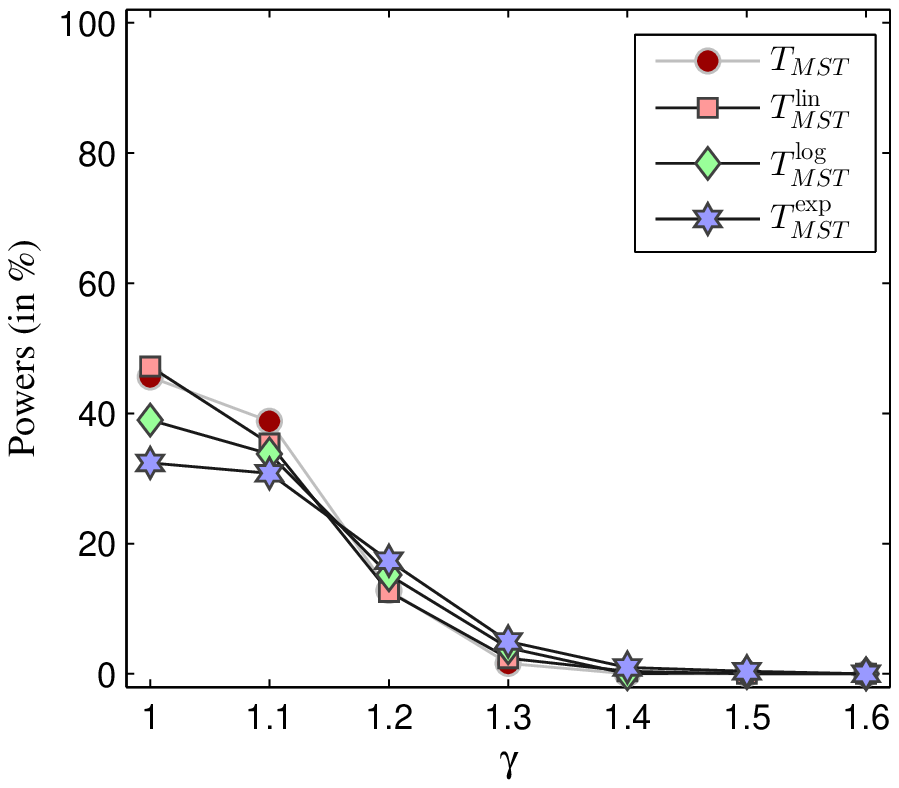}
\end{subfigure}
\begin{subfigure}{0.45\textwidth}
\centering
\subcaption{SHP-run tests}
\includegraphics[height=2.20in,width=\linewidth]{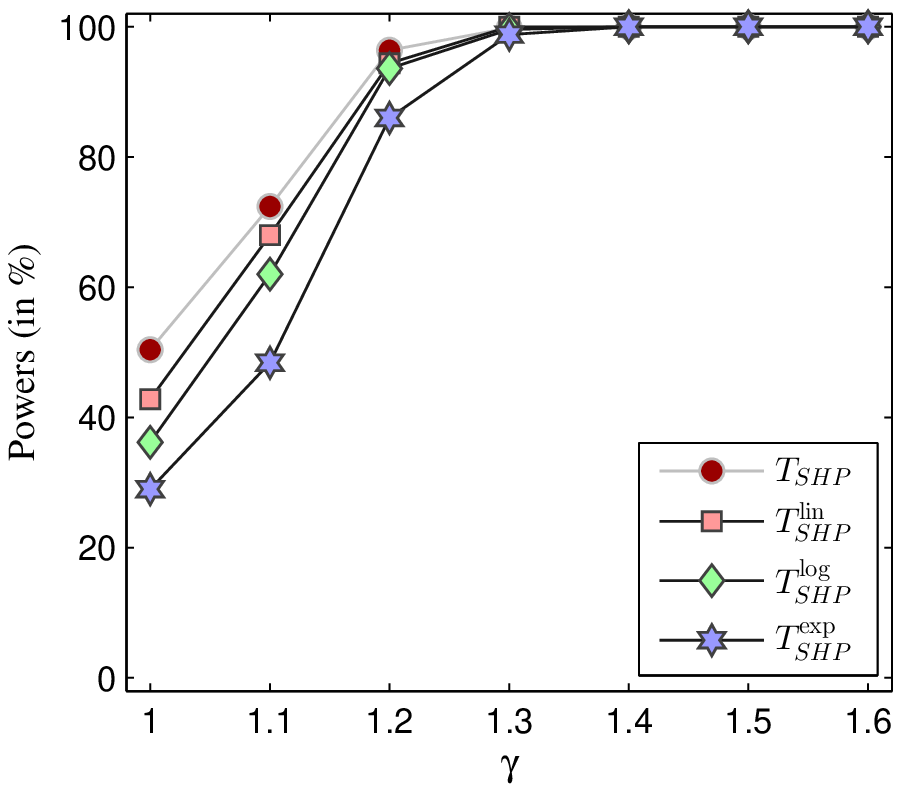}
\end{subfigure}%
\hspace{0.09\textwidth}%
\begin{subfigure}{0.45\textwidth}
\centering
\subcaption{NBP tests}
\includegraphics[height=2.20in,width=\linewidth]{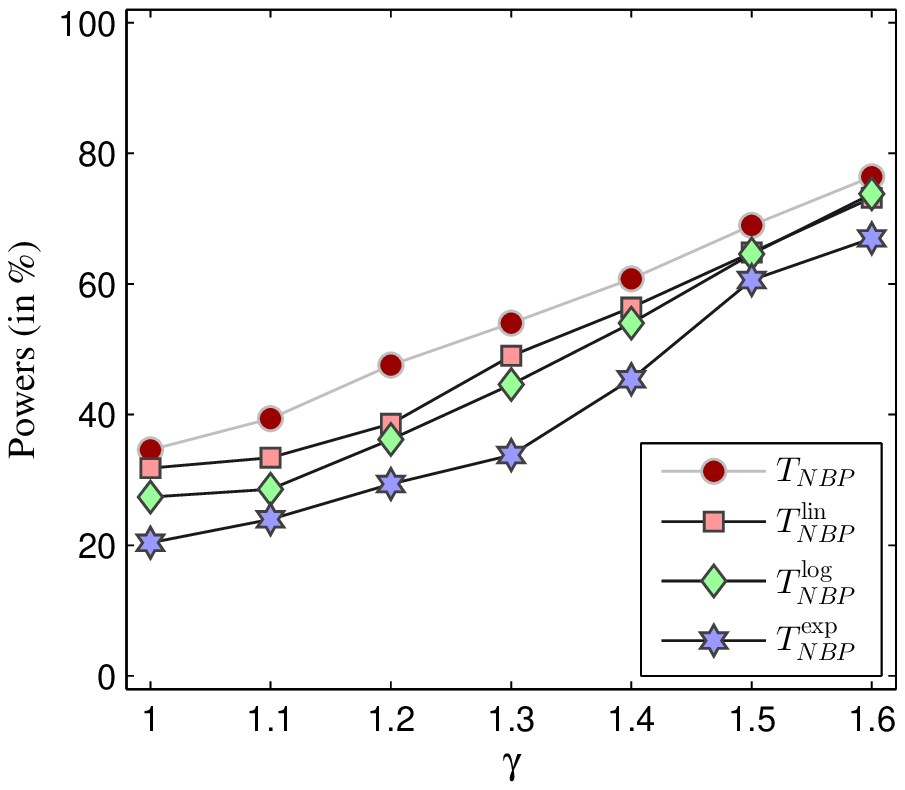}
\end{subfigure}
\caption{Powers of NN, MST-run, SHP-run and NBP tests in Example 3.}
\label{fig:ExA}
\vspace{-0.1in}
\end{figure}

Let us first explain the behavior of the tests based on pairwise Euclidean distances. Consider four independent random vectors $\Xvec_1,\Xvec_2 \sim {\cal N}_d({\bf 0}_d,{\bf I}_d)$ and $\Yvec_1,\Yvec_2 \sim {\cal N}_d(0.2{\bf 1}_d, \gamma^{-1}{\bf I}_d)$.  Here, $d^{-1} \|\Xvec_1 - \Xvec_2\|^2= d^{-1} \sum_{q=1}^d (X_1^{(q)}-X_2^{(q)})^2$, being an average of i.i.d. random variables with finite mean, converges almost surely to $\E(X_1^{(1)}-X_2^{(1)})^2 = 2$ as $d$ diverges to infinity. Similarly, $d^{-1} \|\Yvec_1 - \Yvec_2\|^2$ converges to $2/\gamma$ and $d^{-1} \|\Xvec_1 - \Yvec_1\|^2$ converges to $1 + 1/\gamma+ 0.04$ almost surely as $d$ tends to infinity. Note that similar convergence of pairwise distances can occur even when the measurement variables are neither independent nor identically distributed. In that case, we need some additional conditions to have law of large numbers. Here we give some sufficient conditions in this regard.

\begin{assumption}\label{assumption1}
For $\Wvec \sim F \text{ or } G$, fourth moments of the $W^{(q)}$'s are uniformly bounded.
\end{assumption}
\begin{assumption}\label{assumption2}
Let $\Xvec_{1},\Xvec_{2} \sim F$ and $\Yvec_{1},\Yvec_{2} \sim G$ be independent random vectors. For $\Wvec=\Xvec_{1}-\Xvec_{2}$, $\Yvec_{1}-\Yvec_{2}$ and $\Xvec_{1}-\Yvec_{1}$, $\sum_{1 \le q \ne q^\prime \le d} \corr\big\{(W^{(q)})^{2},(W^{(q^\prime)})^{2}\big\}$ is of the order ${\bf o}(d^{2})$.
\end{assumption}
\begin{assumption}\label{assumption3}
There exist non-negative constants $\nu^2$, $\sigma_{F}^{2}$ and $\sigma_{G}^{2}$ such that $d^{-1}\|\E(\Xvec)-\E(\Yvec)\|^{2}\rightarrow \nu^{2}$, $d^{-1}\sum_{q=1}^{d} \var(X^{(q)})$ $\rightarrow \sigma_{F}^{2}$ and $d^{-1}\sum_{q=1}^{d} \var(Y^{(q)}) \rightarrow \sigma_{G}^{2}$
as $d \to \infty$.
\end{assumption}

Assumption~\ref{assumption2} imposes a form of weak dependence among the measurement variables. It holds for sequence data or stochastic processes with the $\rho$-mixing property. If the measurement variables are i.i.d., then Assumptions~\ref{assumption2} and \ref{assumption3} hold trivially, and instead of Assumption~\ref{assumption1}, we only need the existence of second order moments for the convergence of pairwise Euclidean distances. Assumptions~\ref{assumption1}--\ref{assumption3} are quite common in the HDLSS literature \citep[see, e.g.,][] {HMN05,JM09,BMG14,DSG16}. Assumptions~\ref{assumption1} and \ref{assumption2} ensure that $d^{-1}\big|\|\Wvec\|^{2}-\E\|\Wvec\|^{2}\big|$ converges to zero in probability as $d$ tends to infinity. Now, depending on whether $\Wvec=\Xvec_1-\Xvec_2$, $\Yvec_1-\Yvec_2$ or $\Xvec_1-\Yvec_1$, the limiting value of $d^{-1}\E\|\Wvec\|^{2}$ is obtained from Assumption~\ref{assumption3}. All these facts lead to the following lemma.

\begin{lemma}
\label{lemma:conv_Euclid}
Suppose that $\Xvec_1,\Xvec_2 \sim F$ and $\Yvec_1,\Yvec_2 \sim G$ are independent random vectors. If $F$ and $G$ satisfy Assumptions~\ref{assumption1}--\ref{assumption3}, then as $d$ tends to infinity, $d^{-1/2} \|\Xvec_1-\Xvec_2\|$, $d^{-1/2} \|\Yvec_1-\Yvec_2\|$ and $d^{-1/2} \|\Xvec_1-\Yvec_1\|$ converge in probability to $\sigma_F \sqrt{2}$, $\sigma_G \sqrt{2}$ and ${({\sigma_F^2 + \sigma_G^2 + \nu^2})}^{1/2}$, respectively.
\end{lemma}

Under Assumptions~\ref{assumption1}--\ref{assumption3}, \cite{BMG14} proved the high-dimensional consistency (i.e., the convergence of power to $1$ as $d$ tends to infinity) of the SHP-run test when $\nu^2 > 0$ or $\sigma_F^2 \ne \sigma_G^2$. Under the same condition, one can show this consistency for the NBP test as well (follows using arguments similar to those used in the proof of part $(b)$ of Theorem~\ref{thm:SHP+NBP_consistent}). When $\nu^2 > |\sigma_F^2-\sigma_G^2|$, such high-dimensional consistency can also be proved for the NN test (follows using arguments similar to those used in the proof of part $(a)$ of Theorem~\ref{thm:NN+MST_consistent}) and the MST-run test \citep[see][]{BMG14}. In Example 3, we have $\nu^2=0.04$, $\sigma_F^2=1$ and $\sigma_G^2=\gamma^{-1}$. So, the SHP-run test and the NBP test turn out to be consistent for all values of $\gamma$. That is why these two tests performed well in this example. However, the condition $\nu^2 > |\sigma_F^2 - \sigma_G^2|$ is violated for $\gamma>1.05$. For all higher values of $\gamma$, we have $\nu^2 < \sigma_F^2 - \sigma_G^2$, and as a result, $\Pr\big(\|\Yvec_1-\Yvec_2\| < \|\Xvec_1-\Yvec_1\| < \|\Xvec_1-\Xvec_2\|\big) \rightarrow 1$ as $d \rightarrow \infty$. So, all observations from $G$ have their nearest-neighbors from $G$ with high probability. But, with probability tending to one, all observations from $F$ have their nearest-neighbors from $G$ as well. This violation of neighborhood structure had adverse effects on the performance of  NN and MST-run tests. It is easy to see that when $\nu^2 < \sigma_F^2 - \sigma_G^2$, for any $k<\min\{m,n\}$, $T_{NN} \stackrel{\Pr}{\rightarrow}n/N$ as $d \rightarrow \infty$. This limiting value is close to the mean of  $T_{NN}$ under ${\cal H}_0$ when $m=n$. Also, in such cases, during the construction of the MST of ${\cal G}$,  first a spanning sub-tree on $n$ vertices corresponding to $n$ observations from $G$ is formed. In each of the subsequent steps, an observation from $F$ gets connected to an observation from $G$ {\citep[see][]{BMG14}}. So, as $d \rightarrow \infty$,  $T_{MST} \stackrel{\Pr}{\rightarrow} m+1$, which is equal to its  mean under ${\cal H}_0$ when $m=n$. Therefore, both $T_{NN}$ and $T_{MST}$ fail to cross the corresponding cutoffs. This was the reason behind the poor performance of  NN and MST-run tests in Example 3. Unlike this example, in Examples~\ref{example1} and \ref{example2}, we had $\nu^2=0$ and $\sigma_F^2=\sigma_G^2$. So, $d^{-1/2}\|\Xvec_1-\Xvec_2\|$, $d^{-1/2}\|\Yvec_1-\Yvec_2\|$ and $d^{-1/2}\|\Xvec_1-\Yvec_1\|$ all converged to the same value. Therefore, pairwise Euclidean distances failed to capture the difference between two underlying distributions. As a result, all four tests based on pairwise Euclidean distances had poor results in those examples.

Next, we carry out a theoretical investigation on the high-dimensional behavior of the tests based on $\varphi_{h,\psi}$. For this investigation, we make the following assumption.
\begin{assumption}
\label{assumption4}
Let $\Xvec_1,\Xvec_2 \sim F$, $\Yvec_1,\Yvec_2 \sim G$ be independent random vectors. For $\Wvec=\Xvec_1-\Xvec_2$, $\Yvec_1-\Yvec_2$ and $\Xvec_1-\Yvec_1$, $d^{-1} \sum_{q=1}^d \big\{\psi(|W^{(q)}|) - \E\psi(|W^{(q)}|)\big\} \overset{\Pr}{\rightarrow} 0$ as $d \to \infty$.
\end{assumption}

Assumption~\ref{assumption4} can be viewed as a generalization of Assumptions~\ref{assumption1} and \ref{assumption2}. It holds under Assumptions~\ref{assumption1} and \ref{assumption2} with $(W^{(q)})^2$ replaced by $\psi(|W^{(q)}|)$ (note that Assummption~\ref{assumption1} holds trivially if $\psi$ is bounded). However, it holds in many other situations as well. For instance, \cite{A88} and \cite{dJ95} derived some sufficient conditions based on mixingales. If $h$ is uniformly continuous, under Assumption~\ref{assumption4},  $\big\{\varphi_{h,\psi}(\Xvec_1,\Xvec_2) - \varphi_{h,\psi}^\ast(F,F)\big\}$, $\big\{\varphi_{h,\psi}(\Yvec_1,\Yvec_2) - \varphi_{h,\psi}^\ast(G,G)\big\}$ and $\big\{\varphi_{h,\psi}(\Xvec_1,\Yvec_1) - \varphi_{h,\psi}^\ast(F,G)\big\}$ converge in probability to $0$ as $d$ tends to infinity, where
\begin{align}
\label{eqn:varphistar}
& \varphi^\ast_{h,\psi}(F,F) = h\Big\{{d}^{-1} \sum_{q=1}^d \E\psi(|X_1^{(q)}-X_2^{(q)}|)\Big\},~~ \varphi^\ast_{h,\psi}(G,G) = h\Big\{{d}^{-1}\sum_{q=1}^d \E\psi(|Y_1^{(q)}-Y_2^{(q)}|)\Big\}
\nonumber\\
&~~~~~~~~~~~~~~~~~~~~~~~~~~~~~~~~~~~~~~\text{ and } \varphi^\ast_{h,\psi}(F,G) = h\Big\{{d}^{-1}\sum_{q=1}^d \E\psi(|X_1^{(q)}-Y_1^{(q)}|)\Big\}.
\end{align}
Throughout this article, unless otherwise mentioned, we will assume $h$ to be uniformly continuous. An interesting lemma involving the above-mentioned three quantities is given below.

\begin{lemma}
\label{lemma:separation}
Suppose that $h$ is a strictly increasing, concave function and $\psi^\prime(t)/t$ is a non-constant, monotone function. Then, $e_{h,\psi}(F,G) = 2 \varphi^\ast_{h,\psi}(F,G) - \varphi^\ast_{h,\psi}(F,F) - \varphi^\ast_{h,\psi}(G,G) \ge 0$ for any fixed $d$, and the equality holds if and only if $F$ and $G$ have the same univariate marginal distributions.
\end{lemma}

The quantity $e_{h,\psi}(F,G)$ can be viewed as an energy distance between $F$ and $G$ \citep[see, e.g.,][]{SR04,AZ05}, and it serves as a measure of separation between the two distributions. Lemma~\ref{lemma:separation} shows that for every $d\ge 1$, $e_{h,\psi}(F,G)$ is positive unless the univariate marginals of $F$ and $G$ are identical. Thus, it is reasonable to assume that ${\widetilde e}_{h,\psi}(F,G) = \liminf_{d \to \infty} e_{h,\psi}(F,G)$ $>0$. The following theorem shows the high-dimensional consistency of SHP-run and NBP tests based on $\varphi_{h, \psi}$ under this assumption.

\begin{theorem}
\label{thm:SHP+NBP_consistent}
Let $\Xvec_1,\ldots,\Xvec_m\sim F$ and $\Yvec_1,\ldots,\Yvec_n\sim G$ be independent random vectors, where $F$ and $G$ satisfy Assumption \ref{assumption4} with ${\widetilde e}_{h,\psi}(F,G) = \liminf_{d \to \infty} e_{h,\psi}(F,G) > 0$.\\
$(a)$ If $N/\binom{N}{m} < \alpha$, then the power of the SHP-run test (of level $\alpha$) based on $\varphi_{h,\psi}$  converges to $1$ as $d$ tends to infinity.\\
$(b)$ If $c(m,n) < \alpha$, then the power of the NBP test (of level $\alpha$) based on $\varphi_{h,\psi}$ converges to $1$ as $d$ tends to infinity. Here $c(m,n)$ is given by
$$c(m,n) = \begin{cases}
		 \frac{\left(N/2\right)!}{\binom{N}{m} \left(m/2\right)! \left(n/2\right)!}, &\text{ if both } m,n \text{ are even}\\
		 \frac{2\left(N/2\right)!}{\binom{N}{m} \left((m-1)/2\right)! \left((n-1)/2\right)!}, &\text{ if both } m,n \text{ are odd}\\
		 \frac{\left((N-1)/2\right)!}{\binom{N-1}{m} \left(m/2\right)! \left((n-1)/2\right)!}, &\text{ if } m \text{ is even and }n \text{ is odd}\\
		 \frac{\left((N-1)/2\right)!}{\binom{N-1}{m-1} \left((m-1)/2\right)! \left(n/2\right)!}, &\text{ if } m \text{ is odd and }n \text{ is even}
		\end{cases}
		 $$
\end{theorem}
\vspace{0.1in}

Theorem~\ref{thm:SHP+NBP_consistent} shows that if ${\widetilde e}_{h,\psi}(F,G)>0$, then SHP-run and NBP tests based on $\varphi_{h, \psi}$ have the high-dimensional consistency if the sample sizes are not too small (with $\alpha=0.05$, we need $m,n \ge 5$ and $m,n \ge 8$ for these two tests, respectively). In view of Lemma \ref{lemma:separation}, for our three choices of $h$ and $\psi$, we have ${\widetilde e}_{h,\psi}(F,G)>0$ in Examples 1--3. This was the reason behind the excellent performance by these tests. However, for the tests based on the Euclidean distance (i.e., where $h(t)=\sqrt{t}$ and $\psi(t)=t^2)$, we have ${\widetilde e}_{h,\psi}(F,G)=2(\sigma_F^2+\sigma_G^2+\nu^2)^{1/2}-\sigma_F\sqrt{2}-\sigma_G\sqrt{2}$ (follows from Lemma \ref{lemma:conv_Euclid}), which is positive if and only if $\nu^2 > 0$ or $\sigma_F^2 \ne \sigma_G^2$. This condition was satisfied in Example 3, but not in Examples~\ref{example1} and \ref{example2}. This explains their behavior observed in Figures~\ref{fig:Ex1}--\ref{fig:ExA}.

For the high-dimensional consistency of NN and MST-run tests based on $\varphi_{h,\psi}$, we need some additional conditions, as shown by the following theorem.

\begin{theorem}
\label{thm:NN+MST_consistent}
Let $\Xvec_1,\ldots,\Xvec_m \sim F$ and $\Yvec_1,\ldots,\Yvec_n\sim G$ be independent random vectors, where $F$ and $G$ satisfy Assumption \ref{assumption4}. Also assume that both $\liminf_{d \to \infty} \{\varphi_{h,\psi}^{\ast}(F,G) - \varphi_{h,\psi}^\ast(F,F)\}$ and $\liminf_{d \to \infty} \{\varphi_{h,\psi}^\ast(F,G) - \varphi_{h,\psi}^\ast(G,G)\}$ are positive.\\
$(a)$ Define $N_0 = \lceil N/(k+1) \rceil$ and $m_0 = \lceil \min\{m,n\}/(k+1) \rceil$ (here $\lceil t \rceil$ denotes the smallest integer larger than or equal to $t$). If $k < \min\{m,n\}$ and $\binom{N_0}{m_0} < \alpha \binom{N}{m}$, then the power of the NN test (of level $\alpha$) based on $\varphi_{h,\psi}$ converges to $1$ as $d$ tends to infinity.\\
$(b)$ If $\max\{\lfloor N/m \rfloor, \lfloor N/n \rfloor\} < \alpha \binom{N}{m}$, then the power of the  MST-run test (of level $\alpha$) based on $\varphi_{h,\psi}$ converges to $1$ as $d$ tends to infinity.
\end{theorem}

The conditions $\liminf_{d \to \infty} \{\varphi_{h,\psi}^{\ast}(F,G) - \varphi_{h,\psi}^\ast(F,F)\} > 0$ and $\liminf_{d \to \infty} \{\varphi_{h,\psi}^\ast(F,G) - \varphi_{h,\psi}^\ast(G,G)\}>0$ {ensure that} the neighborhood structure, in terms of $\varphi_{h,\psi}$, is preserved in high dimensions,  i.e., an observation has its nearest-neighbor from the same distribution with high probability. In Example~\ref{example1}, we have $\lim_{d \to \infty} \varphi_{h,\psi}^\ast(F,F) = \lim_{d \to \infty} \varphi^\ast_{h,\psi}(G,G)$ (it is clear from the descriptions of the two classes). So, in view of Lemma~\ref{lemma:separation}, both $\lim_{d \to \infty} \{\varphi_{h,\psi}^{\ast}(F,G) - \varphi_{h,\psi}^\ast(F,F)\}$ and $\lim_{d \to \infty} \{\varphi_{h,\psi}^\ast(F,G) - \varphi_{h,\psi}^\ast(G,G)\}$ are positive. But that is not the case in Examples~\ref{example2}~and~3, where $\varphi_{h,\psi}^\ast(F,G)$ lies between $\varphi_{h,\psi}^\ast(F,F)$ and $\varphi_{h,\psi}^\ast(G,G)$ for all $d$. Because of this violation of neighborhood structure, NN and MST-run tests based on $\varphi_{h,\psi}$ had such poor results. The following theorem shows that in such situations, powers of these two tests may even drop down to zero.

\begin{theorem}
\label{thm:NN+MST_inconsistent}
Let $\Xvec_1,\ldots,\Xvec_m\sim F$ and $\Yvec_1,\ldots,\Yvec_n\sim G$ be independent random vectors, where $F$ and $G$ satisfy Assumption \ref{assumption4}. Also assume that $\limsup_{d \to \infty} \{\varphi_{h,\psi}^\ast(F,G) - \varphi_{h,\psi}^\ast(F,F)\} < 0$ (interchange $F$ and $G$ if required, and in that case, interchange $m$ and $n$, accordingly).\\
$(a)$ If $k < \min\{m,n\}$ and $(m-1)/n > (1+\alpha)/(1-\alpha)$, then the power of the NN test (of level $\alpha$) based on $\varphi_{h,\psi}$ converges to $0$ as $d$ tends to infinity.\\
$(b)$ If $m/n > (1+\alpha)/(1-\alpha)$, then the power of the MST-run test (of level $\alpha$) based on $\varphi_{h,\psi}$ converges to $0$ as $d$ tends to infinity.
\end{theorem}

Note that the conditions involving $m$ and $n$ in Theorem~\ref{thm:NN+MST_inconsistent} are only sufficient. These conditions do not hold in Example 3, but NN and MST-run tests based on $\varphi_{h,\psi}$ had powers close to $0$. To overcome these limitations of NN and MST-run tests, in the next section, we introduce a new class of dissimilarity measures and use it to modify NN and MST-run tests.

\section{Modified NN and MST-run tests based on a new class of dissimilarity indices}
\label{sec:MADD}

Given the combined sample ${\cal Z}_N$, we define the dissimilarity index between two observations $\xvec$ and $\yvec$ in ${\cal Z}_N$ as
\begin{equation}
\rho_{h,\psi}(\xvec,\yvec) = \frac{1}{N-2} \sum_{\zvec \in {\cal Z}_N \setminus \{\xvec,\yvec\}} \Bigl|\varphi_{h,\psi}(\xvec,\zvec) - \varphi_{h,\psi}(\yvec,\zvec)\Bigr|,
\end{equation}
where $\varphi_{h,\psi}$ is as defined in Section~\ref{sec:intro}. Since this dissimilarity index is based on the {\bf M}ean of {\bf A}bsolute {\bf D}ifferences of pairwise {\bf D}istances, we call it {\MADD}. Using $h(t)=\sqrt{t}$ and $\psi(t) = t^2$, we get MADD based on the Euclidean distance. This is given by
\begin{equation}\label{eq:MADD}
\rho_0(\xvec,\yvec) = \frac{1}{N-2} \sum_{\zvec \in {\cal Z}_N \setminus \{\xvec,\yvec\}} d^{-1/2}\big|\|\xvec-\zvec\|-\|\yvec-\zvec\|\big|.
\end{equation}
Note that the Euclidean distance usually increases with the dimension at the rate of $d^{1/2}$ \citep[see, e.g.,][]{HMN05}. This justifies the use of $d^{-1/2}$ as the scaling factor. MADD has several desirable properties as a dissimilarity index. One such property is mentioned below.
\begin{lemma}
\label{lemma:semi-metric}
For $N\ge 3$, the dissimilarity index $\rho_{h,\psi}$ is a semi-metric on ${\cal Z}_N$.
\end{lemma}

The index $\rho_{h,\psi}$ is not a metric since $\rho_{h,\psi}(\xvec,\yvec)=0$ does not necessarily imply $\xvec=\yvec$. However, if $F$ and $G$ are absolutely continuous, then for any $\xvec \ne \yvec$, $\rho_{h,\psi}(\xvec,\yvec)$ is strictly positive with probability $1$. So, $\rho_{h,\psi}$ behaves like a metric for all practical purposes. When $\varphi_{h,\psi}$ is a metric, using the triangle inequality, we also get $\rho_{h,\psi}(\xvec,\yvec)\le \varphi_{h,\psi}(\xvec,\yvec)$. So, closeness in terms of $\varphi_{h,\psi}$ indicates closeness in terms of $\rho_{h,\psi}$, but not the other way around. For instance, in the case of high-dimensional data, unlike the Euclidean distance, $\rho_0$ usually takes small values for observations
from the same distribution. This is shown by the following lemma.

\begin{lemma}
\label{lemma:conv_MADD}
Suppose that $\Xvec_1,\Xvec_2 \sim F$ and $\Yvec_1,\Yvec_2 \sim G$ are independent random vectors. If $F$ and $G$ satisfy Assumptions~\ref{assumption1}--\ref{assumption3}, then as $d$ tends to infinity, $\rho_0(\Xvec_1,\Xvec_2)$ and $\rho_0(\Yvec_1,\Yvec_2)$ converge in probability to $0$, while $\rho_0(\Xvec_1,\Yvec_1)$ converges in probability to a non-negative quantity $\widetilde{\rho}_0(F,G)$, which takes the value $0$ if and only if $\nu^2 = 0$ and $\sigma_F^2 = \sigma_G^2$.
\end{lemma}

Therefore, if $\nu^2>0$ or $\sigma_F^2 \ne \sigma_G^2$, all observations have their neighbors (in terms of $\rho_0$) from their own distributions with high probability. Because of this phenomenon, tests based on $\rho_0$ outperform those based on the Euclidean distance in a wide variety of high-dimensional problems. In this context, we have the following result.

\begin{theorem}
\label{thm:rho0_consistency}
Suppose that $\Xvec_1,\ldots,\Xvec_m \sim F$ and $\Yvec_1,\ldots,\Yvec_n \sim G$ are independent random vectors, where $F$ and $G$ satisfy Assumptions~\ref{assumption1}--\ref{assumption3} with $\nu^2>0$ or $\sigma_F^2 \neq \sigma_G^2$.\\
$(a)$ Define $N_0$ and $m_0$ as in \autoref{thm:NN+MST_consistent}. If $k < \min\{m,n\}$ and $\binom{N_0}{m_0} < \alpha \binom{N}{m}$, then the power of the NN test (of level $\alpha$) based on ${\rho_0}$ converges to $1$ as $d$ tends to infinity.\\
$(b)$ If $\max\{\lfloor N/m \rfloor, \lfloor N/n \rfloor\} < \alpha \binom{N}{m}$, then the power of the MST-run test (of level $\alpha$) based on ${\rho_0}$ converges to $1$ as $d$ tends to infinity.
\end{theorem}

This theorem shows that if the sample sizes are not too small, NN and MST-run tests based on $\rho_0$ have the high-dimensional consistency. For $\alpha = 0.05$, the conditions $\binom{N_0}{m_0} < \alpha \binom{N}{m}$ and $\max\{\lfloor N/m \rfloor, \lfloor N/n \rfloor\}< \alpha \binom{N}{m}$ hold whenever $m,n \ge 4$. NN and MST-run tests based on the Euclidean distance have this consistency when $\nu^2>|\sigma_F^2-\sigma_G^2|$ \citep[see, e.g.,][]{BMG14,MBG15}. In Example 3, we had $\nu^2 > 0$ and $\sigma_F^2\neq \sigma_G^2$, but $\nu^2$ was smaller than $|\sigma_F^2-\sigma_G^2|$. So, while the tests based on the Euclidean distance had powers close to zero, those based on $\rho_0$ had excellent performance (see the cureves corresponding to ${\widetilde T}_{NN}$ and ${\widetilde T}_{MST}$ in Figure~\ref{fig:ExA_MADD}). But that was not the case in Examples 1 and 2
(see Figures~\ref{fig:Ex1_MADD}~and~\ref{fig:Ex2_MADD}), where we had $\sigma_F^2=\sigma_G^2$ and $\nu^2=0$. In those examples, NN and MST-run tests based on $\rho_0$ also had poor performance. In such cases, we need to use tests based on other versions of MADD. In this article, we use MADD based on three other choices of $h$ and $\psi$: $(i)$ $h(t)=t, \psi(t)=t$, $(ii)$ $h(t)=t, \psi(t)=\log(1+t)$ and $(iii)$ $h(t)=t,\psi(t)=1-\exp(-t)$, and the corresponding dissimilarity indices are denoted by $\rho_1$, $\rho_2$ and $\rho_3$, respectively. Figures~\ref{fig:Ex1_MADD}~and~\ref{fig:Ex2_MADD} show that NN and MST-run tests based on these three indices had excellent performance in Examples 1 and 2 (see the curves corresponding to ${\widetilde T}^{\rm lin}_{NN}$, ${\widetilde T}^{\log}_{NN}$, ${\widetilde T}^{\exp}_{NN}$ and ${\widetilde T}^{\rm lin}_{MST}$, ${\widetilde T}^{\log}_{MST}$, ${\widetilde T}^{\exp}_{MST}$, respectively). They had good performance in Example 3 as well (see Figure~\ref{fig:ExA_MADD}). In that example, the difference between the two distributions was only in their locations and scales. So, the test based on $\rho_0$ had slightly higher powers than these tests.

\begin{figure}[t]
\begin{subfigure}{0.45\textwidth}
\centering
\subcaption{Tests based on nearest-neighbors}
\includegraphics[height=2.20in,width=\linewidth]{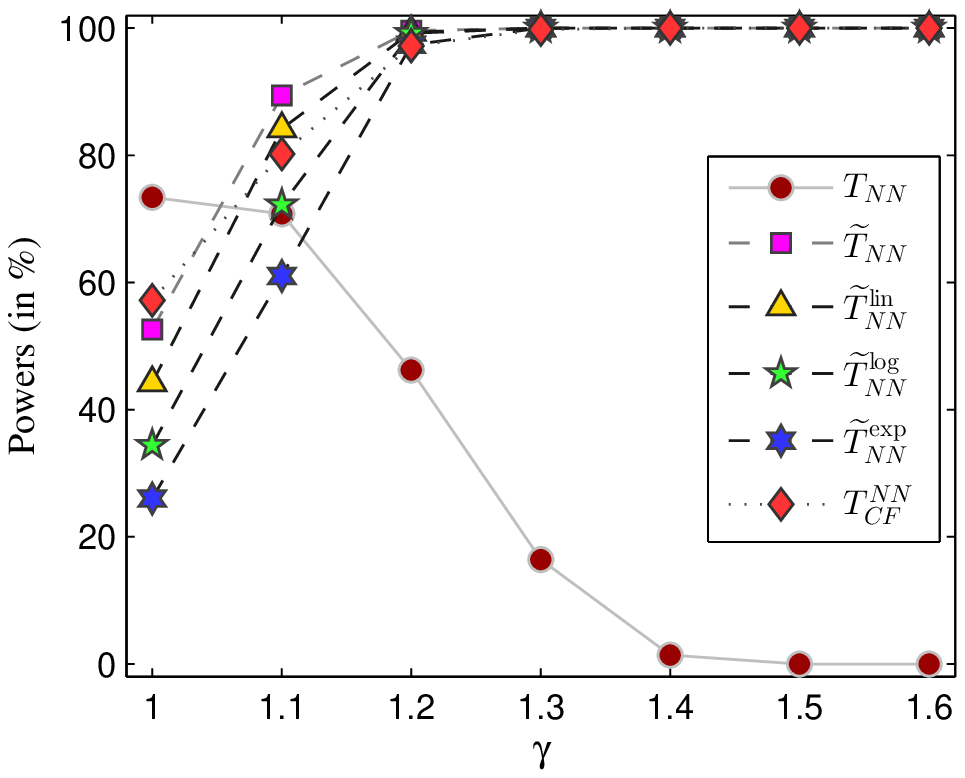}
\end{subfigure}%
\hspace{0.09\textwidth}%
\begin{subfigure}{0.45\textwidth}
\centering
\subcaption{Tests based on MST}
\includegraphics[height=2.20in,width=\linewidth]{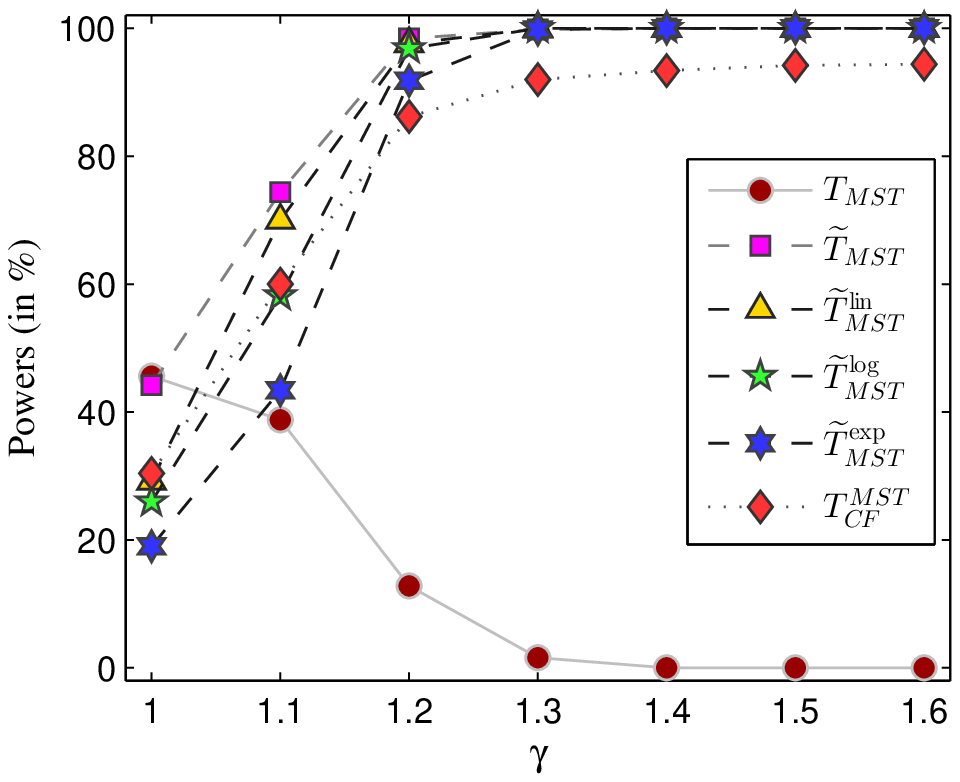}
\end{subfigure}
\caption{Powers of different tests in Example 3.}
\label{fig:ExA_MADD}
\vspace{-0.2in}
\end{figure}

\begin{figure}[!b]
\begin{subfigure}{0.45\textwidth}
\centering
\subcaption{Tests based on nearest-neighbors}
\includegraphics[height=2.20in,width=\linewidth]{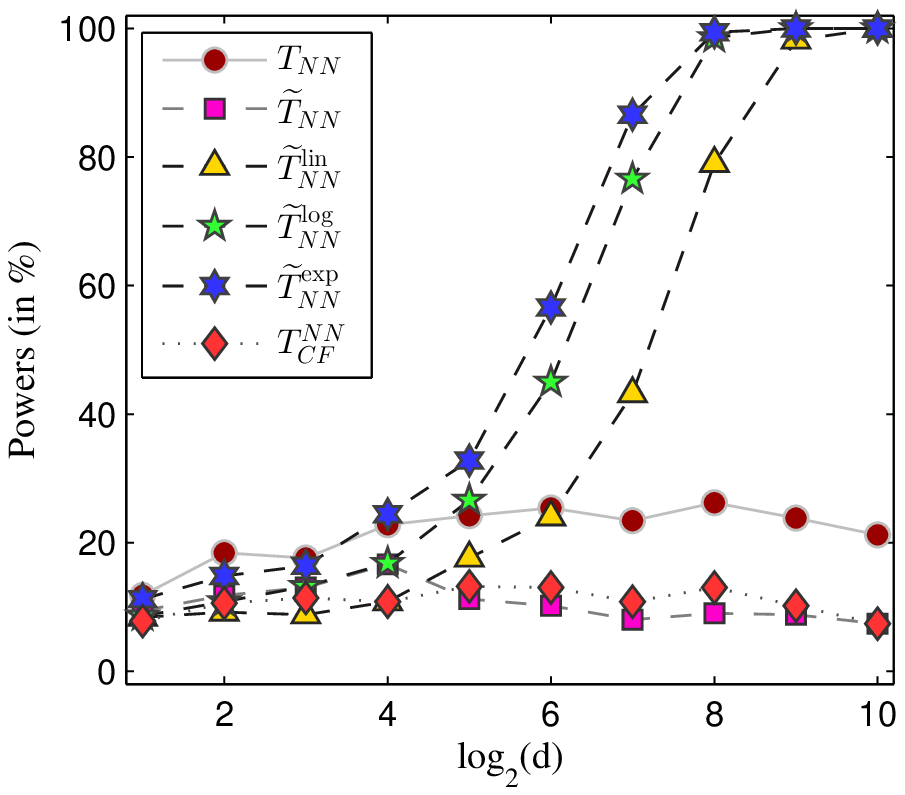}
\end{subfigure}%
\hspace{0.09\textwidth}%
\begin{subfigure}{0.45\textwidth}
\centering
\subcaption{Tests based on MST}
\includegraphics[height=2.20in,width=\linewidth]{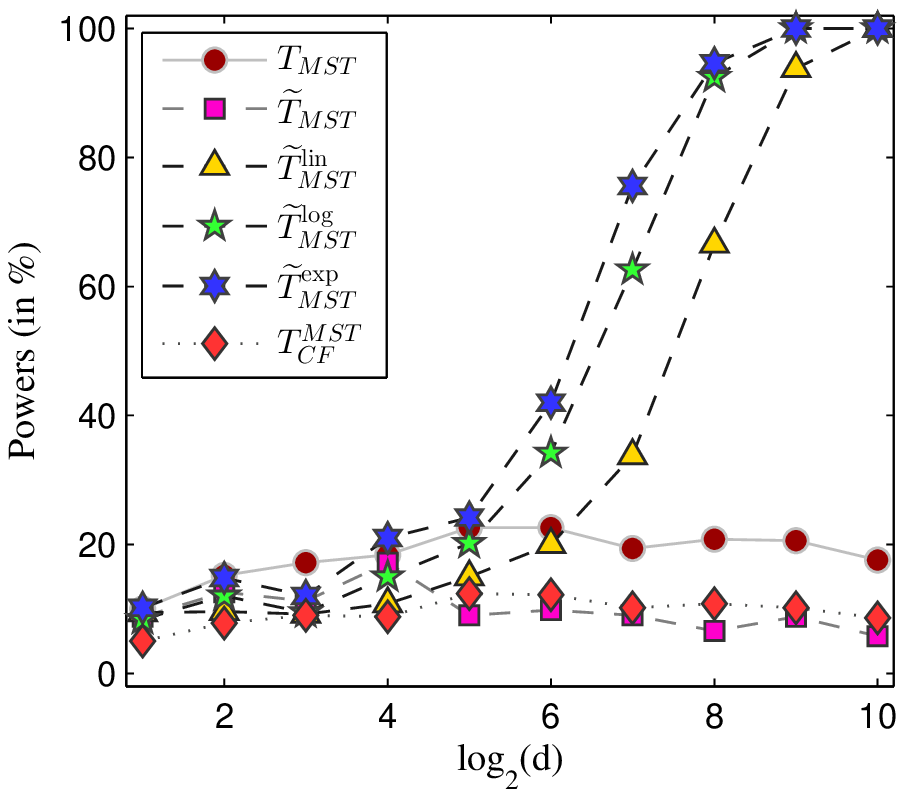}
\end{subfigure}
\caption{Powers of different tests in Example \ref{example1}.}
\label{fig:Ex1_MADD}
\end{figure}

\begin{figure}[h]
\begin{subfigure}{0.45\textwidth}
\centering
\subcaption{Tests based on nearest-neighbors}
\includegraphics[height=2.20in,width=\linewidth]{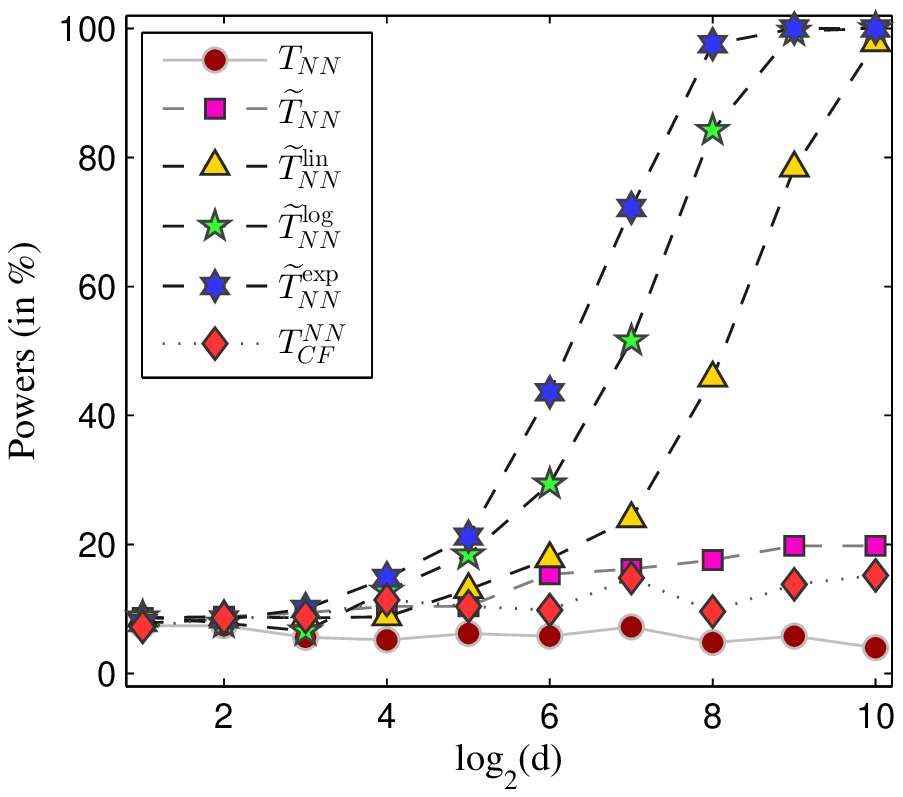}
\end{subfigure}%
\hspace{0.09\textwidth}%
\begin{subfigure}{0.45\textwidth}
\centering
\subcaption{Tests based on MST}
\includegraphics[height=2.20in,width=\linewidth]{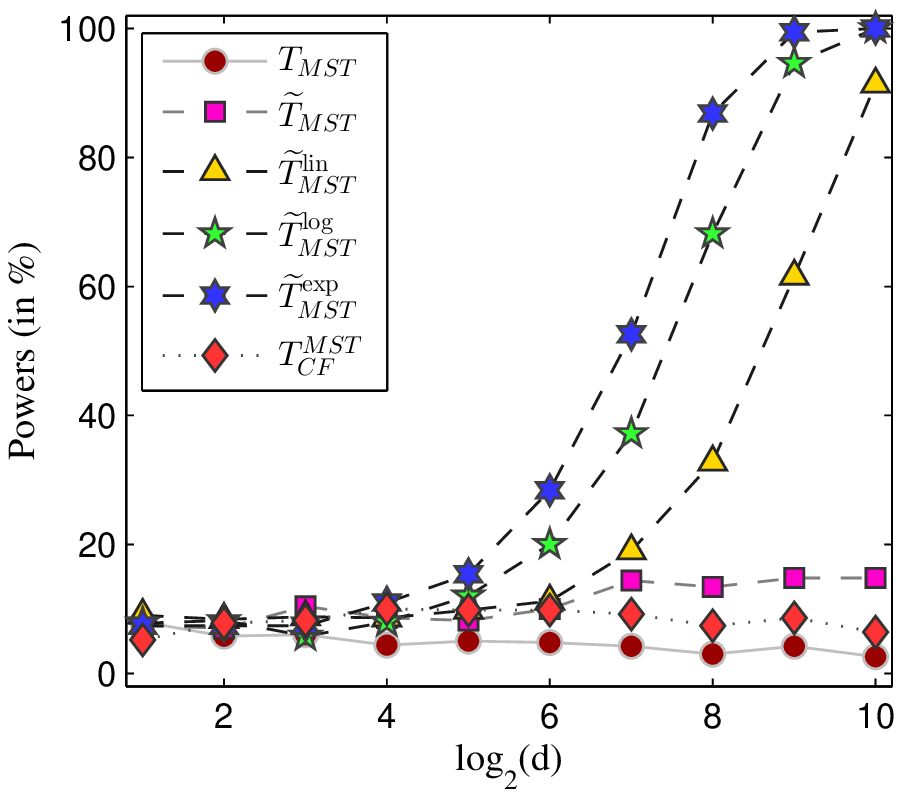}
\end{subfigure}
\caption{Powers of different tests in Example \ref{example2}.}
\label{fig:Ex2_MADD}
\end{figure}

Recently, \cite{CF17} developed a general framework to construct graph-based two-sample tests for multivariate data, where one counts the numbers of $\Xvec\Xvec$-type and $\Yvec\Yvec$-type edges ($S_{xx}$ and $S_{yy}$, say) in a sub-graph of ${\cal G}$ and compute the deviations from their expected values under ${\cal H}_0$. The test statistic is defined as $T_{CF}={(\Svec - \E_{{\cal H}_0}(\Svec))}^{\top}\big[\var_{{\cal H}_0}(\Svec)\big]^{-1}(\Svec - \E_{{\cal H}_0}(\Svec))$, where $\Svec=(S_{xx},~ S_{yy})^{\top}$. In particular, \cite{CF17} used $k$-nearest-neighbor graph (k-NN graph) and MST of ${\cal G}$ for all numerical work. The $k$-NN graph is an undirected sub-graph of ${\cal G}$, which contains the edge $(\uvec,\vvec)$ if either $\vvec$ is among the $k$ nearest-neighbors of $\uvec$ or $\uvec$ is among the $k$ nearest neighbors of $\vvec$. These two tests (henceforth, referred to as CF~-~NN and CF-MST tests, respectively) perform better than the usual NN and MST-run tests based on the Euclidean distance in many examples \citep[see][]{CF17}. Throughout this article, we use $k=3$ for the CF-NN test to have fair comparison with the NN test. These tests worked well in Example 3 (see the curves corresponding to $T_{CF}^{NN}$ and $T_{CF}^{MST}$ in Figure~\ref{fig:ExA_MADD}). But just like the tests based on $\rho_0$, they had poor performance in Examples~\ref{example1} and \ref{example2} (see Figures~\ref{fig:Ex1_MADD}~and~\ref{fig:Ex2_MADD}).

Note that since $h$ is uniformly continuous, under Assumption~\ref{assumption4}, we have the probability convergence of $\big|\varphi_{h,\psi}(\Xvec_1,\Xvec_2)-\varphi_{h,\psi}^*(F,F)\big|$, $\big|\varphi_{h,\psi}(\Yvec_1,\Yvec_2)-\varphi_{h,\psi}^*(G,G)\big|$ and
$\big|\varphi_{h,\psi}(\Xvec_1,\Yvec_1)-\varphi_{h,\psi}^*(F,G)\big|$ to $0$ as $d$ tends to infinity. This leads to the probability convergence of $\rho_{h,\psi}(\Xvec_1,\Xvec_2)$, $\rho_{h,\psi}(\Yvec_1,\Yvec_2)$ and $\{\rho_{h,\psi}(\Xvec_1,\Yvec_1)-\rho_{h,\psi}^{*}(F,G)\}$ to $0$, where $\rho_{h,\psi}^{*}(F,G)=(N-2)^{-1}\{(m-1)$ $|\varphi_{h,\psi}^*(F,G)-\varphi_{h,\psi}^*(F,F)| + (n-1)|\varphi_{h,\psi}^*(F,G)-\varphi_{h,\psi}^*(G,G)|\} \ge 0$. But, in order to preserve the neighborhood structure (in terms of $\rho_{h,\psi}$) in high dimensions, we need to choose $h$ and $\psi$  so that $\rho_{h,\psi}^{*}(F,G)$ is strictly positive. The following lemma provides some guidance in this regard.
\begin{lemma}
\label{lemma:MADD_general}
Let $h,\psi:[0,\infty) \rightarrow [0,\infty)$ be strictly increasing functions such that $h(0)=\psi(0)=0$ and $\psi^\prime(t)/t$ is a non-constant, monotone function. Then, for every $d \ge 1$, $\rho_{h,\psi}^\ast(F,G)$ is positive unless $F$ and $G$ have the same univariate marginal distributions.
\end{lemma}
In view of Lemma~\ref{lemma:MADD_general}, it is reasonable to make the following assumption.
\begin{assumption}\label{assumption5}
$\widetilde{\rho}_{h,\psi}(F,G) = \liminf_{d \to \infty} \rho_{h,\psi}^{\ast}(F,G) > 0$.
\end{assumption}
In the proof of Lemma~\ref{lemma:MADD_general}, one can see that for any fixed $d$, $\rho_{h,\psi}^\ast(F,G)= 0$ if and only if $e_{F,G}^{(q)}=2 \E\psi(|X_1^{(q)}-Y_1^{(q)}|) - \E\psi(|X_1^{(q)}-X_2^{(q)}|)- \E\psi(|Y_1^{(q)}-Y_2^{(q)}|)=0$ for $q=1,\ldots,d$. This quantity $e_{F,G}^{(q)}$ is an energy distance between the $q$-th univariate marginals of $F$ and $G$ {\citep[see, e.g.,][]{SR13}} that gives signal against ${\cal H}_0$. Now, $\widetilde{\rho}_{h,\psi}(F,G)$ becomes $0$ only when $\liminf_{d \to \infty} d^{-1}\sum_{q=1}^{d} e_{F,G}^{(q)}=0$. So, Assumption~\ref{assumption5} asserts that the average signal is asymptotically non-negligible. In classical asymptotic regime, we consider $d$ to be fixed and expect to get more information as $m$ and $n$ increase. But, in the HDLSS asymptotic regime, where we consider $m$ and $n$ to be fixed, we expect to get more information as $d$ increases. This is ensured by Assumptions~\ref{assumption4} and \ref{assumption5}. The following theorem shows the high-dimensional consistency of modified NN  and MST-run tests based on $\rho_{h, \psi}$  under these assumptions.
\begin{theorem}
\label{thm:Power_MADDgeneral}
Suppose that $\Xvec_1,\ldots,\Xvec_m \sim F$ and $\Yvec_1,\ldots,\Yvec_n \sim G$ are independent random vectors, and $\rho_{h,\psi}$ is used to construct the test statistics, where $h$ and $\psi$ satisfy the conditions of Lemma~\ref{lemma:MADD_general}. Then, under Assumptions~\ref{assumption4} and \ref{assumption5}, we get the following results.\\
$(a)$ Let $N_0$ and $m_0$ be defined as in Theorem 4. If $k < \min\{m,n\}$ and $\binom{N_0}{m_0} <\alpha \binom{N}{m}$, then the power of the NN test (of level $\alpha$) based on $\rho_{h,\psi}$ converges to $1$  as $d$ tends to infinity.\\
$(b)$ If $\max\{\lfloor N/m \rfloor, \lfloor N/n \rfloor\} < \alpha \binom{N}{m}$, then the power of the MST-run test (of level $\alpha$) based on $\rho_{h,\psi}$ converges to $1$  as $d$ tends to infinity.
\end{theorem}

The function $\psi(t)=t^2$ used to define $\rho_0$ does not satisfy the conditions stated in Lemma~\ref{lemma:MADD_general}, but the choices of $h$ and $\psi$ used for $\rho_1$, $\rho_2$ and $\rho_3$ satisfy them. For these three choices of $h$ and $\psi$, ${\widetilde \rho}_{h,\psi}(F,G)$ turns out to be positive in Examples~\ref{example1} and \ref{example2}. This was the reason behind the excellent performance by the tests based on these three indices in those two examples, where the tests based on $\rho_0$ had performed poorly.

\subsection{Performance under weak signal}
\label{sec:weak_signal}
In Theorem~\ref{thm:Power_MADDgeneral}, we have established the consistency of the tests based on $\rho_{h,\psi}$ when $ {\widetilde \rho}_{h,\psi}(F,G) > 0$, or equivalently,  $\liminf_{d \to \infty} d^{-1} \sum_{q=1}^d e_{FG}^{(q)} >0$. So, we need $\sum_{q=1}^d e_{FG}^{(q)}$, the total signal against ${\cal H}_0$, to increase at least at the rate of $d$. But if only a few of the measurement variables carry information against ${\cal H}_0$, we may have $\liminf_{d \to \infty} d^{-1} \sum_{q=1}^d e_{FG}^{(q)} =0$. Next, we investigate the high-dimensional behavior of the tests based on $\rho_{h,\psi}$ in such situations. For two independent random vectors $\Uvec,\Vvec \sim F$ or $G$, let us assume that $\var\{\sum_{q=1}^d \psi(|U^{(q)}-V^{(q)}|)\} = {\bf O}(\vartheta^2(d))$. If the measurement variables are highly correlated, we usually have $\vartheta^2(d)={\bf O}(d^2)$. But weak dependence among the measurement variables leads to $\vartheta^2(d)={\bf o}(d^2)$. For instance, when they are $m$-dependent, one gets $\vartheta^2(d) = d$. Now, for our investigation, we make the following assumption, which is weaker than Assumption~\ref{assumption5}.
\begin{assumption}\label{assumption6}
As $d$ tends to infinity, $\rho^\ast_{h,\psi}(F,G) ~d/\vartheta(d)$ diverges to infinity.
\end{assumption}
In Assumption~\ref{assumption6}, we allow $\rho^\ast_{h,\psi}(F,G)$ to converge to $0$, but at a rate slower than that of $\vartheta(d)/d$. For instance, for an $m$-dependent sequence, we allow $\rho^\ast_{h,\psi}(F,G)$ to converge to $0$ at a rate slower than $d^{-1/2}$. Even when the measurement variables are not $m$-dependent, under certain weak dependence assumptions on the underlying distributions, we have $\vartheta^2(d) = d L(d)$, where $L$ is a slowly varying function \citep[see][Chap. 2]{LL96}. In that case, we allow $\rho^\ast_{h,\psi}(F,G)$  to converge to $0$ at a rate slower than $d^{-1/2} L^{1/2}(d)$. Under Assumption~\ref{assumption6}, $\rho_{h,\psi}$ preserves the neighborhood structure in high dimensions when $h$ is Lipschitz continuous, and the high-dimensional consistency of the resulting tests follows from that. The result is stated below.
\begin{theorem}
\label{thm:MADD_weaker}
Suppose that $\Xvec_1,\ldots,\Xvec_m \sim F$ and $\Yvec_1,\ldots,\Yvec_n \sim G$ are independent random vectors, where $F$ and $G$ satisfy Assumptions~\ref{assumption4} and \ref{assumption6}. If $h$ is Lipschitz continuous, then
$$
\Pr\Big[\min_{i,j} \rho_{h,\psi}(\Xvec_i,\Yvec_j) > \max\big\{\max_{i \ne j} \rho_{h,\psi}(\Xvec_i,\Xvec_j) , \max_{i \ne j} \rho_{h,\psi}(\Yvec_i,\Yvec_j)\big\}\Big] \to 1 \text{ as } d \to \infty.
$$
Consequently, if $m$ and $n$ satisfy the conditions of Theorem~\ref{thm:Power_MADDgeneral}, then the powers of NN and MST-run tests (of level $\alpha$) based on $\rho_{h,\psi}$ converge to $1$ as $d$ tends to infinity.
\end{theorem}

For $\Xvec \sim F$ and $\Yvec \sim G$, under the assumptions of Theorem~\ref{thm:MADD_weaker}, we get $\rho_{h,\psi}(\Xvec,\Yvec) = \rho^\ast_{h,\psi}(F,G)$ + ${\bf O}_{P}(\vartheta(d)/d)$ (see the proof of Theorem~\ref{thm:MADD_weaker}). Here $\rho^\ast_{h,\psi}(F,G)$ can be viewed as the signal against ${\cal H}_0$, while $\vartheta(d)/d=\big[\var\{d^{-1}\sum_{q=1}^d \psi(|X^{(q)}-Y^{(q)}|)\}\big]^{1/2}$ can be interpreted as stochastic variation or noise. Theorem~\ref{thm:MADD_weaker} shows the high-dimensional consistency of NN and MST-run tests based on $\rho_{h,\psi}$ when $h$ is  Lipschitz continuous and the signal-to-noise ratio diverges with $d$. Similar results can be obtained even when $h$ is not Lipschitz. For instance, in the case of $\rho_0$, $h(t) = \sqrt{t}$ is not Lipschitz continuous, but we have the following result.

\begin{theorem}
\label{thm:Euclid_weaker}
Suppose that $\Xvec_1,\ldots,\Xvec_m \sim F$ and $\Yvec_1,\ldots,\Yvec_n \sim G$ are independent random vectors, where $F$ and $G$ have means $\muvec_F, \muvec_G$ and dispersion matrices $\sigmat_F, \sigmat_G$, respectively. Further assume that $\liminf_{d \to \infty} \min\{tr(\sigmat_F),tr(\sigmat_G)\}/\vartheta(d) > 0$, where $\vartheta^2(d)$ is the order of $\var(\|\Wvec\|^2)$ for $\Wvec=\Xvec_1-\Xvec_2$, $\Yvec_1-\Yvec_2$ and $\Xvec_1-\Yvec_1$. If $\|\muvec_F-\muvec_G\|^2/\vartheta(d)$ or $|tr(\sigmat_F) - tr(\sigmat_G)|/\vartheta(d)$ diverges to infinity as $d$ increases, then
\begin{equation*}
{\Pr}\Big[\min_{i,j} \rho_{0}(\Xvec_i,\Yvec_j) > \max\big\{\max_{i \ne j} \rho_{0}(\Xvec_i,\Xvec_j) , \max_{i \ne j} \rho_{0}(\Yvec_i,\Yvec_j)\big\}\Big] \to 1 \text{ as } d \to \infty.
\end{equation*}
Consequently, if $m$ and $n$ satisfy the conditions of {Theorem~\ref{thm:rho0_consistency}}, then the powers of NN and MST-run tests (of level $\alpha$) based on $\rho_0$ converge to $1$ as $d$ tends to infinity.\end{theorem}

Thus, when the measurement variables are $m$-dependent, for the consistency of the tests based on $\rho_0$, we need either $d^{-1/2} \|\muvec_F - \muvec_G\|^2$ or $d^{-1/2} |tr(\sigmat_F)-tr(\sigmat_G)|$ to diverge to infinity as $d$ increases. This condition is much weaker than the conditions assumed in Theorem 8.

\subsection{Computational issues}
Computation of {\MADD} between two data points has an associated cost of the order ${\bf O}(dn)$ compared to ${\bf O}(d)$ needed for the Euclidean distance or $\varphi_{h,\psi}$. But in the HDLSS set up, where $d$ is much larger than $n$, these are of the same asymptotic order. Moreover, after computing all pairwise distances, the steps used for obtaining the test statistics are the same in all cases. Therefore, for HDLSS data, though the tests based on {\MADD} require more time compared to the corresponding tests based on the Euclidean distance or $\varphi_{h,\psi}$, the time difference is not that significant. This is quite evident from the following table, which shows average computing times required by NN and MST-run tests based on the Euclidean distance and $\rho_0$ for various dimensions and sample sizes. We used MATLAB codes for all these tests, and they were run on a computer with 8 GB RAM, having Intel Core i7 CPU with the clock speed of 2.20GHz.

\setlength{\tabcolsep}{1.5mm}
\begin{table}[h!]
\centering
\small
\caption{Average run times (of $100$ trials) for different tests (in seconds).}
\begin{tabular}{|c|c|c|c|c|c|c|c|c|c|c|c|c|c|}\hline
Distance/ & \multicolumn{6}{|c|}{$m=n=20$} & \multicolumn{6}{c|}{$m=n=40$}\\ \cline{2-13}
Dissimilarity & \multicolumn{2}{|c|}{$d=200$} & \multicolumn{2}{|c|}{$d=500$}& \multicolumn{2}{|c|}{$d=1000$}& \multicolumn{2}{|c|}{$d=200$}& \multicolumn{2}{c|}{$d=500$}& \multicolumn{2}{c|}{$d=1000$}\\ \cline{2-13}
Index& NN & MST & NN & MST& NN & MST& NN & MST& NN & MST& NN & MST\\ \hline
Euclidean & 7.40 & 7.02 &13.22 &13.18& 23.14  &24.60 & 14.86 & 13.10 &27.33 &25.61     &46.20&46.56  \\
MADD ($\rho_0$)& 7.51& 7.45 & 13.35&14.09 & 23.66 &25.67 & 15.45 & 13.77 &28.32 &26.64 &47.03& 47.18\\ \hline \end{tabular}
\end{table}

\vspace{-0.1in}
\section{Results from the analysis of simulated and real data sets}
\label{sec:simulation}

Using Examples~\ref{example1}--3, we have already demonstrated the usefulness of MADD for NN and MST-run tests in high-dimensional set up. In this section, we analyze four more simulated data sets (two involving mixture distributions and two involving weak signals) and two real data sets for further evaluation of these tests. In each of these cases, we repeated the experiment 500 times to compute the powers of different tests, which are shown in Figures~7--12. For all simulated data sets, we used $m=n=20$ as before. Here we also report the results for CF-NN and CF-MST tests to facilitate comparison. Throughout this section, all tests are considered to have $5\%$ nominal level.

\subsection{Analysis of simulated data sets}

Examples~\ref{example5} and \ref{example6} deal with mixture distributions, where at least one of the two population distributions is a mixture of two multivariate distributions with convex supports.

\setcounter{example}{3}
\begin{example}\label{example5}
$F$ is an equal mixture of ${\cal N}_d(0.3{\bf 1}_d,{\bf I}_d)$ and ${\cal N}_d(-0.3{\bf 1}_d,4{\bf I}_d)$, while $G$ is an equal mixture of ${\cal N}_d(0.3\alphavec_d,{\bf I}_d)$ and ${\cal N}_d(-0.3\alphavec_d,4{\bf I}_d)$, where $\alphavec_d = (1,-1,\ldots,(-1)^{d+1})^\top$.
\end{example}

\begin{example}\label{example6}
Let ${\cal C}_{d,r} = \{\xvec \in \real^d : |x^{(q)}| \le r/2 ~ \forall q=1,\ldots,d\}$ be a $d$-dimensional hypercube with sides of length $r$. While $F$ is the uniform distribution on ${\cal C}_{d,1}$, $G$ is an equal mixture of two uniform distributions on ${\cal C}_{d,0.9}$ and ${\cal C}_{d,1.1}$, respectively.
\end{example}

In Examples 4 and 5, NN and MST-run tests based on the Euclidean distance performed poorly. Performances of CF-NN and CF-MST tests were even worse in Example 4. In Example 5, they performed better, but their powers were much lower than those of all MADD based tests (i.e., the tests based on $\rho_0$, $\rho_1$, $\rho_2$ and $\rho_3$) considered here. All MADD based tests had similar powers in Example 5. In Example 4 also, they had competitive performance, while the tests based on $\rho_0$ and $\rho_1$ had an edge. In these two examples, Assumptions~\ref{assumption1}--\ref{assumption4} do not hold for the mixture distributions, but they hold for each component distribution. If we consider each of them as a separate distribution, using the distance concentration phenomenon, we can explain the reasons behind poor performance of the Euclidean distance based tests and superiority of their modified versions based on MADD.

\begin{figure}[h]
\centering
\begin{subfigure}{0.45\textwidth}
\centering
\subcaption{Tests based on nearest-neighbors}
\includegraphics[height=2.20in,width=\linewidth]{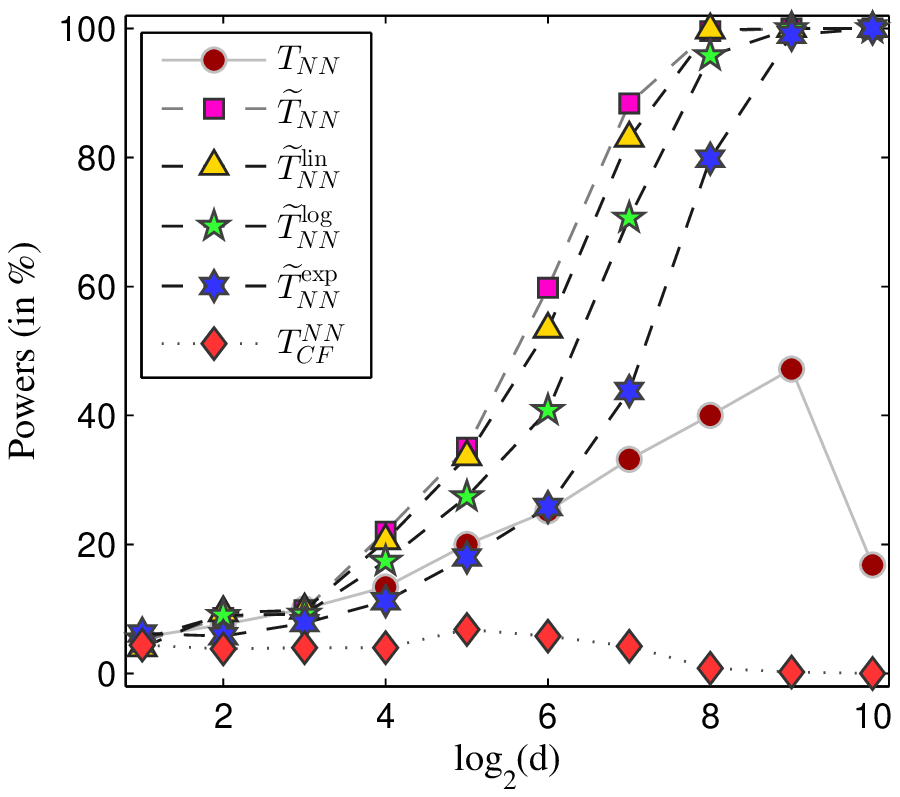}
\end{subfigure}%
\hspace{0.09\textwidth}%
\begin{subfigure}{0.45\textwidth}
\centering
\subcaption{Tests based on MST}
\includegraphics[height=2.20in,width=\linewidth]{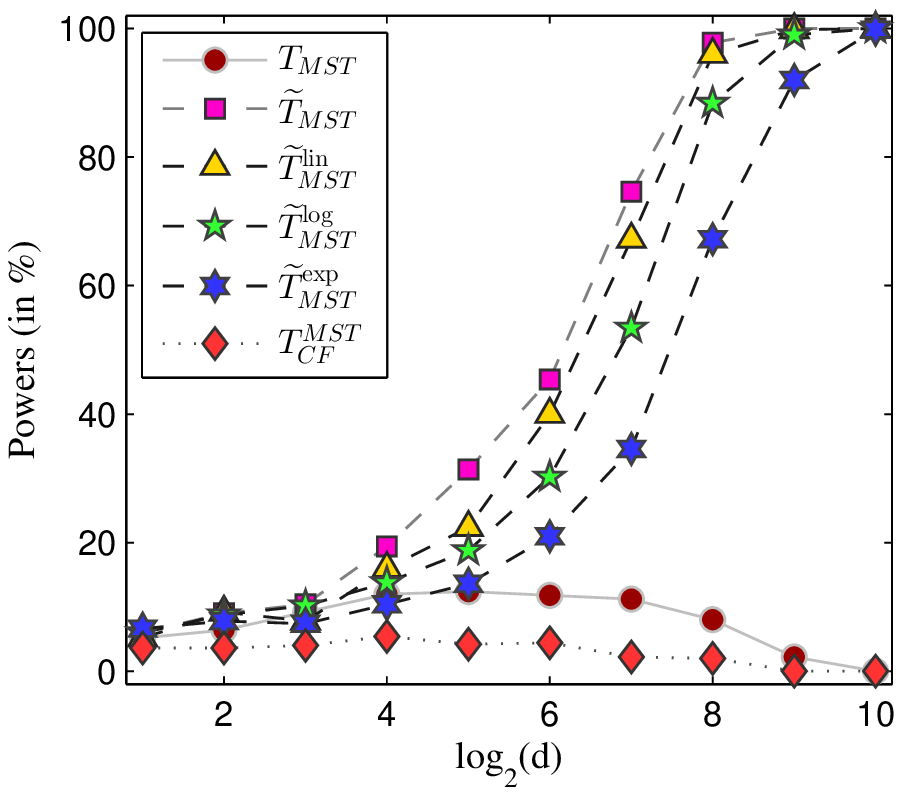}
\end{subfigure}
\caption{Powers of different tests in Example~\ref{example5}.}
\end{figure}

Our next two examples involve alternatives with sparse signals, where only a fraction of the measurement variables contain information against ${\cal H}_0$, and that fraction shrinks to $0$ as the dimension increases. So,  Assumption~\ref{assumption5} does not hold in these examples.

\begin{figure}[h]
\centering
\begin{subfigure}{0.45\textwidth}
\centering
\subcaption{Tests based on nearest-neighbors}
\includegraphics[height=2.20in,width=\linewidth]{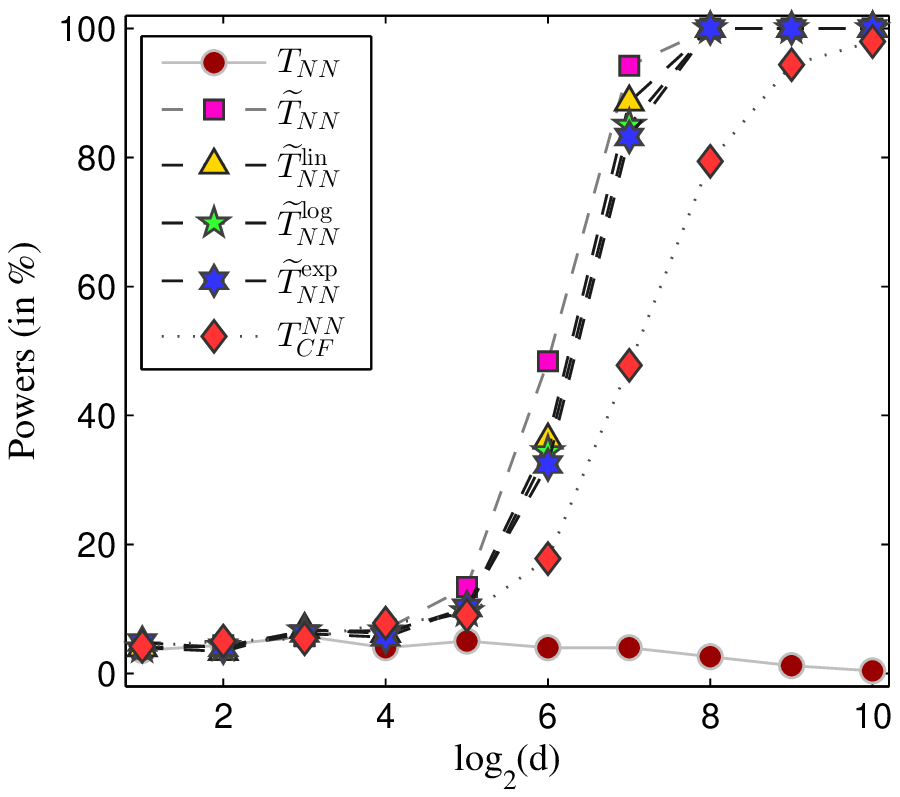}
\end{subfigure}%
\hspace{0.09\textwidth}%
\begin{subfigure}{0.45\textwidth}
\centering
\subcaption{Tests based on MST}
\includegraphics[height=2.20in,width=\linewidth]{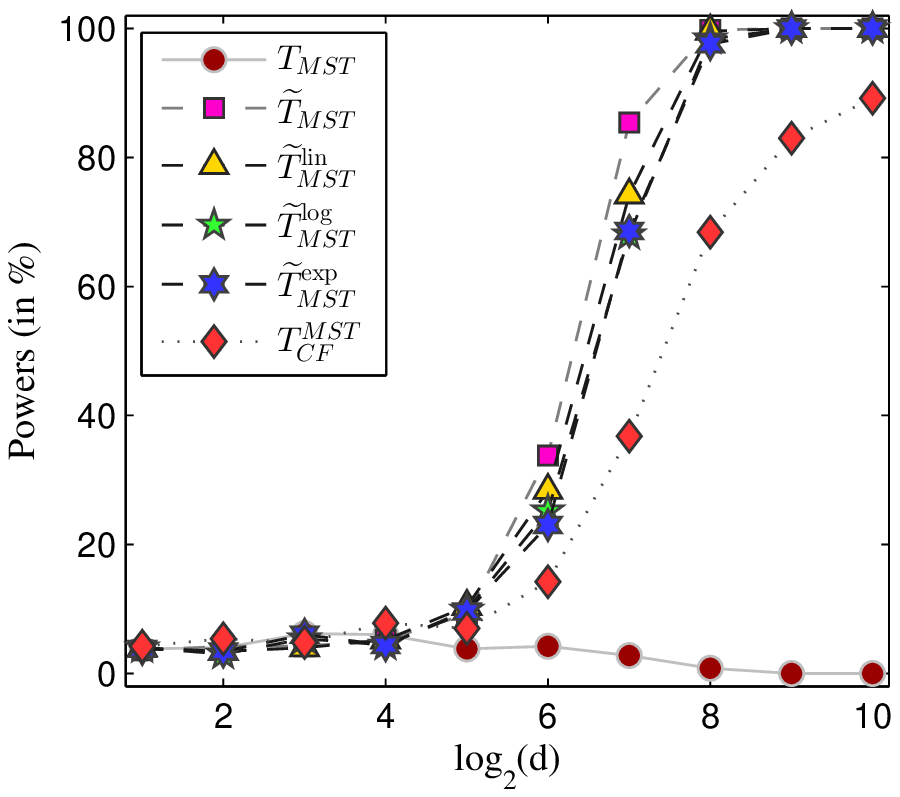}
\end{subfigure}
\caption{Powers of different tests in Example~\ref{example6}.}
\vspace{-0.15in}
\end{figure}

\begin{example}\label{example7}
We consider two distributions ${\cal N}_d({\bf 0}_d,{\bf I}_d)$ and ${\cal N}_d(\muvec_d,\Lambdat_d)$, where $\muvec_d = (\mu_1,\ldots,\mu_d)^\top$ with $\mu_i = \sqrt{0.01\log(d)}$ for $i=1,\ldots,d^{1/2}$ and $0$ otherwise. The diagonal matrix $\Lambdat_d$ has the first $d^{1/2}$ elements equal to $0.5\log(d)$ and the rest equal to $1$. 
\vspace{-0.05in}
\end{example}
In this example, modified tests based on $\rho_0$ outperformed all other tests considered here. Chen and Friedman's tests (CF tests) had the second best performance. Here, the two distributions differ in their locations and scales. So, as expected, tests based on other choices of MADD had slightly lower powers than those based on $\rho_0$. Note that for $\rho_0$, $\sum_{q=1}^{d} e_{F,G}^{(q)}$ is of the order ${\bf O}(d^{1/2} \log(d))$, while for $\rho_3$, it is of the order ${\bf O}(d^{1/2})$ since the function $\psi$ is bounded. NN and  MST-run tests based on the Euclidean distance had powers close to zero throughout.

\begin{figure}[h]
\centering
\begin{subfigure}{0.45\textwidth}
\centering
\subcaption{Tests based on nearest-neighbors}
\includegraphics[height=2.20in,width=\linewidth]{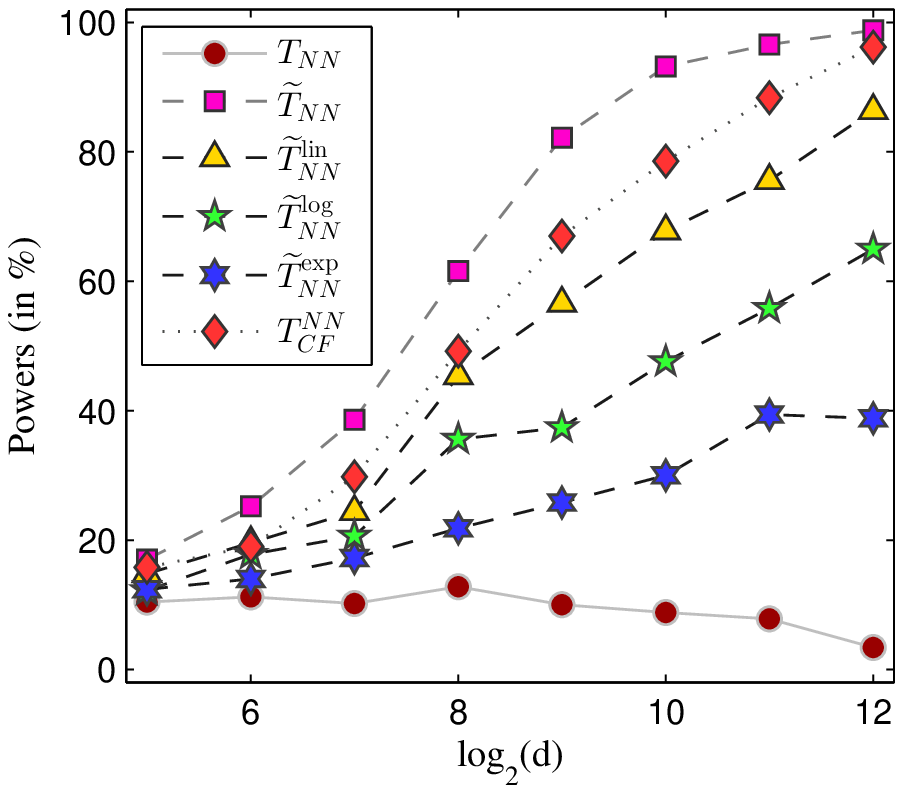}
\end{subfigure}%
\hspace{0.09\textwidth}%
\begin{subfigure}{0.45\textwidth}
\centering
\subcaption{Tests based on MST}
\includegraphics[height=2.20in,width=\linewidth]{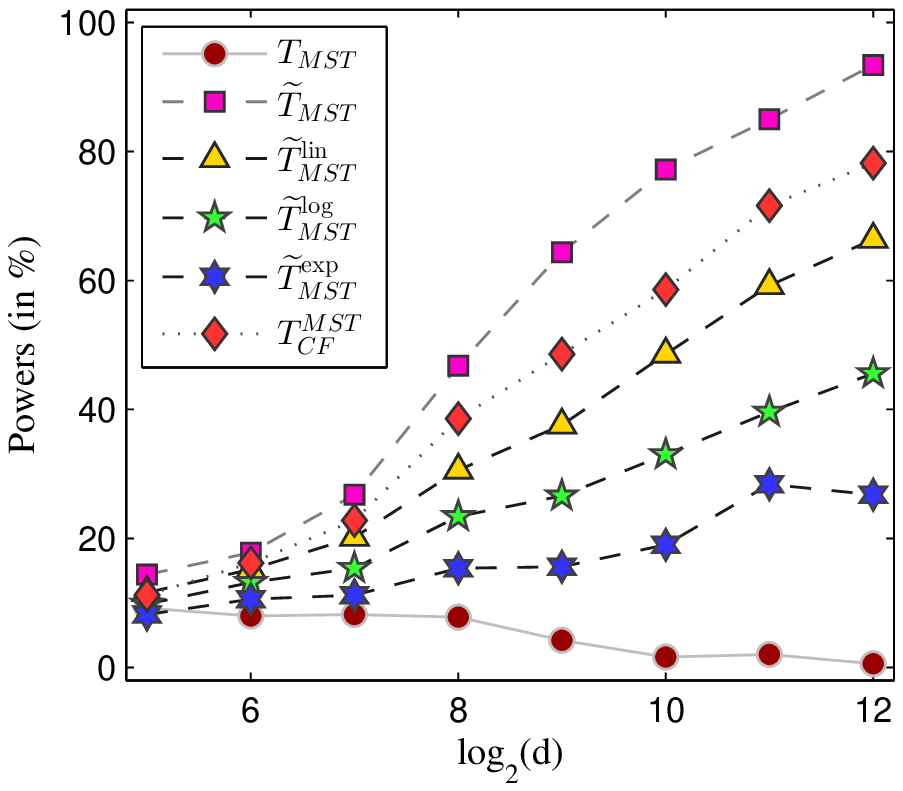}
\end{subfigure}
\caption{Powers of different tests in Example~\ref{example7}.}
\end{figure}

\vspace{-0.15in}
\begin{example}\label{example8}
Both distributions have independent measurement variables. They are distributed as ${\cal N}(0,1)$ for the first population. For the second population, the first $d^{2/3}$ variables are $t_3(0,1/3)$, and the rest are ${\cal N}(0,1)$. So, these two populations have the same location and dispersion structure, while the first $d^{2/3}$ variables differ only in their shapes.
\end{example}

\begin{figure}[t]
\centering
\begin{subfigure}{0.45\textwidth}
\centering
\subcaption{Tests based on nearest-neighbors}
\includegraphics[height=2.20in,width=\linewidth]{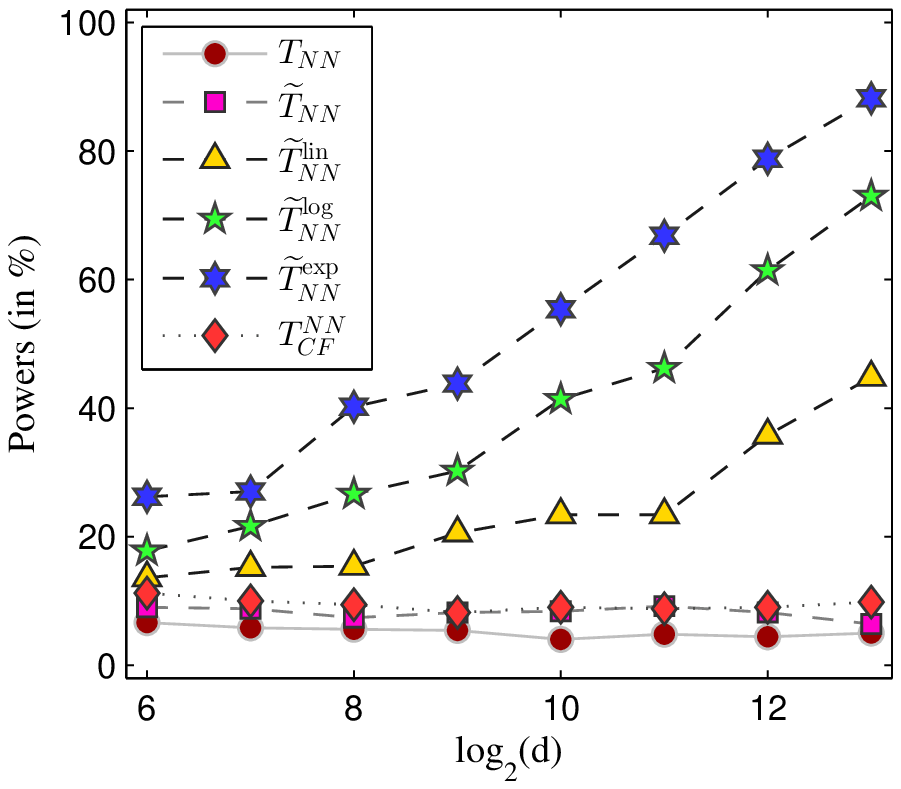}
\end{subfigure}%
\hspace{0.09\textwidth}%
\begin{subfigure}{0.45\textwidth}
\centering
\subcaption{Tests based on MST}
\includegraphics[height=2.20in,width=\linewidth]{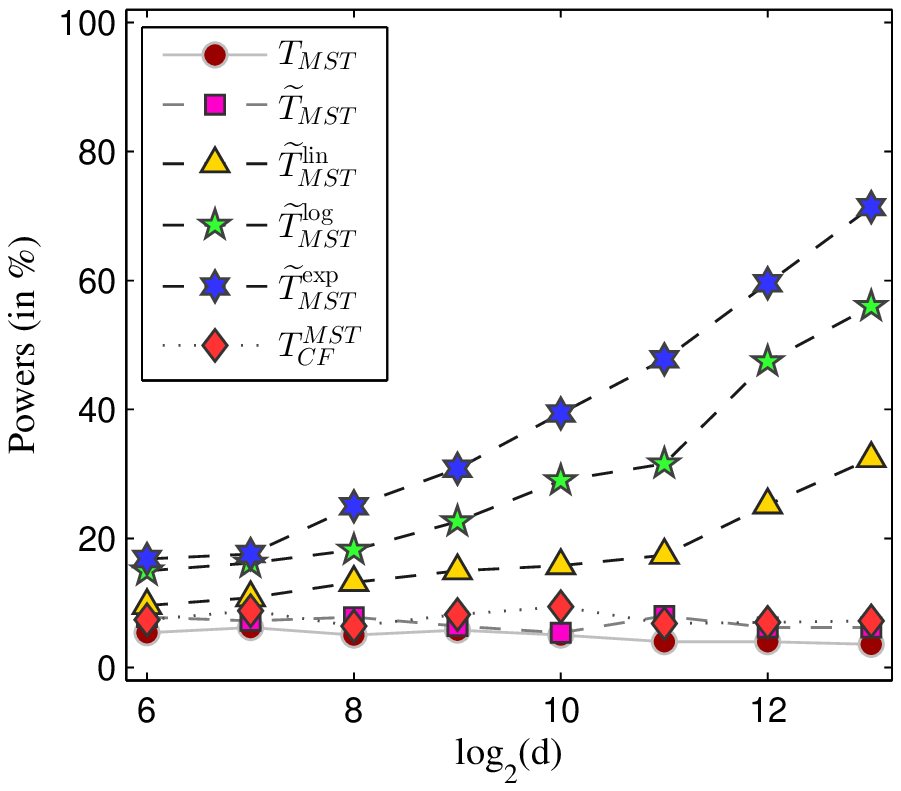}
\end{subfigure}
\caption{Powers of different tests in Example~\ref{example8}.}
\end{figure}
In this example, we observed a different picture. Modified tests based on $\rho_1$, $\rho_2$ and $\rho_3$ performed much better than all other tests considered here. Among these modified tests, the tests based on $\rho_3$ had superior performance. Note that in this example, two distributions have the same location and scale, but they differ in their univariate marginal distributions. In such a case, pairwise Euclidean distances failed to extract the information regarding the separation between two distributions. So, CF tests, the tests based on the Euclidean distance and those based on $\rho_0$, all had powers close to the nominal level.

\subsection{Analysis of benchmark data sets}

We also analyzed two benchmark data sets, the {\sc Gun-Point} data and the {\sc Lighting-2} data, for further evaluation of our proposed tests. These data sets are taken from the {\it UCR Time Series Classification Archive} (\url{http://www.cs.ucr.edu/~eamonn/time_series_data/}). They have been extensively used in the literature of supervised classification. In both of these data sets, we have reasonable separation between the two distributions. So, assuming ${\cal H}_0$ to be false, we compared different tests based on their powers. These data sets consist of separate training and test samples. For our analysis, we merged these sets and following \cite{BMG14}, we used random subsamples of different sizes from the whole data set keeping the proportions of observations from different distributions as close as they are in the original data set. Each experiment was repeated 500 times to compute the powers of different tests, and they are shown in Figures 11 and 12.

\begin{figure}[h]
\centering
\begin{subfigure}{0.45\textwidth}
\centering
\subcaption{Tests based on nearest-neighbors}
\includegraphics[height=2.20in,width=\linewidth]{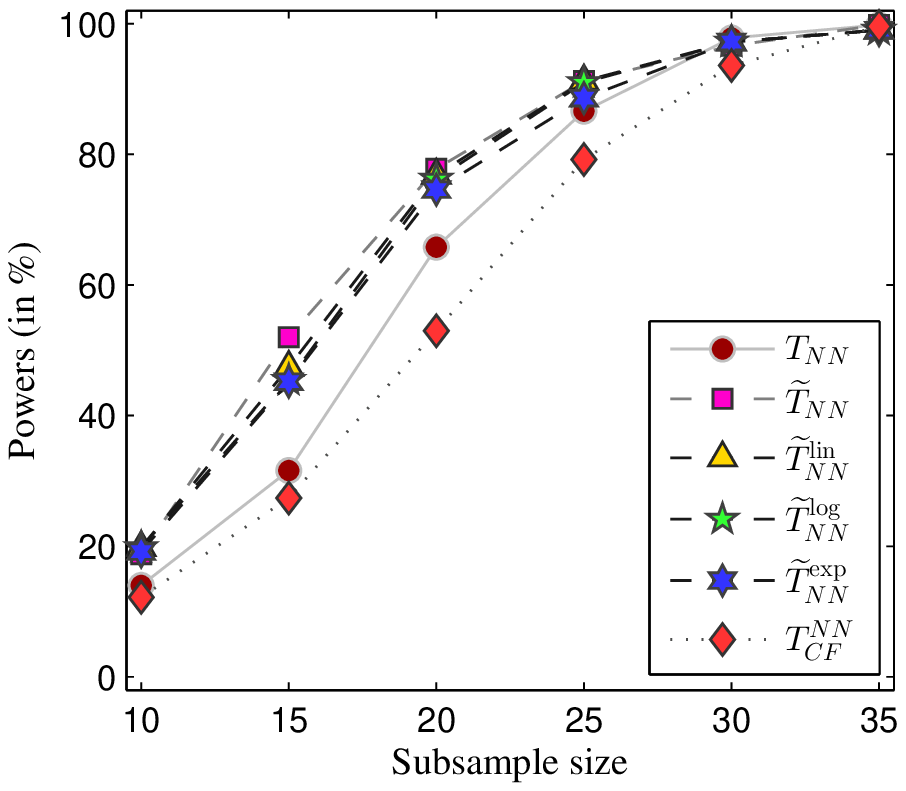}
\end{subfigure}%
\hspace{0.09\textwidth}%
\begin{subfigure}{0.45\textwidth}
\centering
\subcaption{Tests based on MST}
\includegraphics[height=2.20in,width=\linewidth]{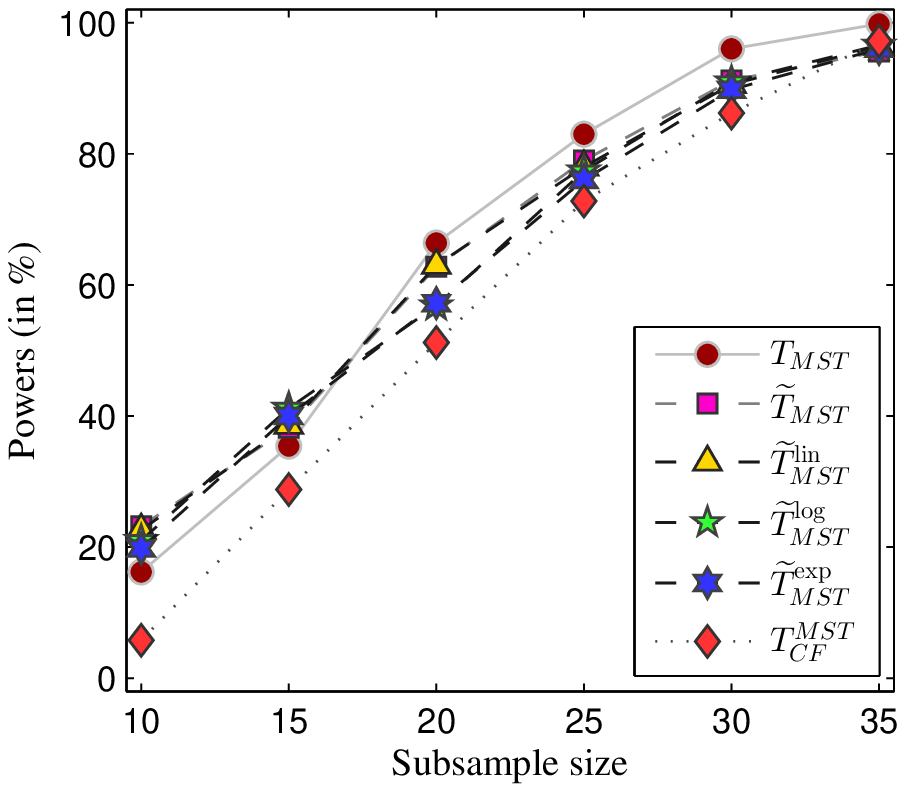}
\end{subfigure}
\caption{Powers of different tests in {\sc Gun-Point} data.}
\end{figure}

\begin{figure}[h]
\centering
\begin{subfigure}{0.45\textwidth}
\centering
\subcaption{Tests based on nearest-neighbors}
\includegraphics[height=2.20in,width=\linewidth]{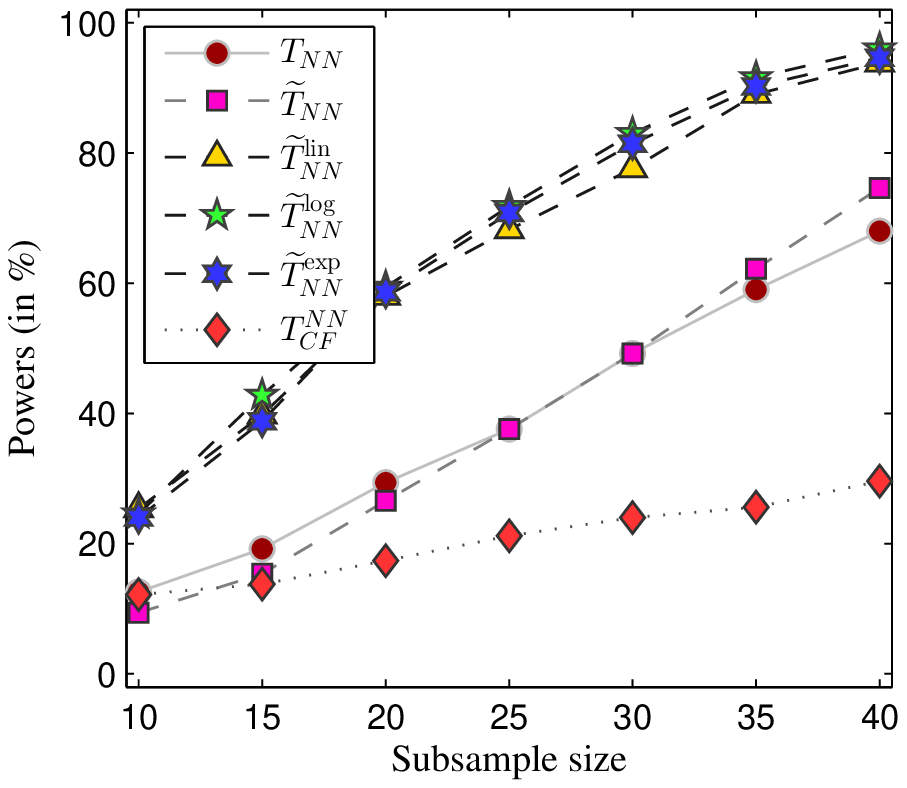}
\end{subfigure}%
\hspace{0.09\textwidth}%
\begin{subfigure}{0.45\textwidth}
\centering
\subcaption{Tests based on MST}
\includegraphics[height=2.20in,width=\linewidth]{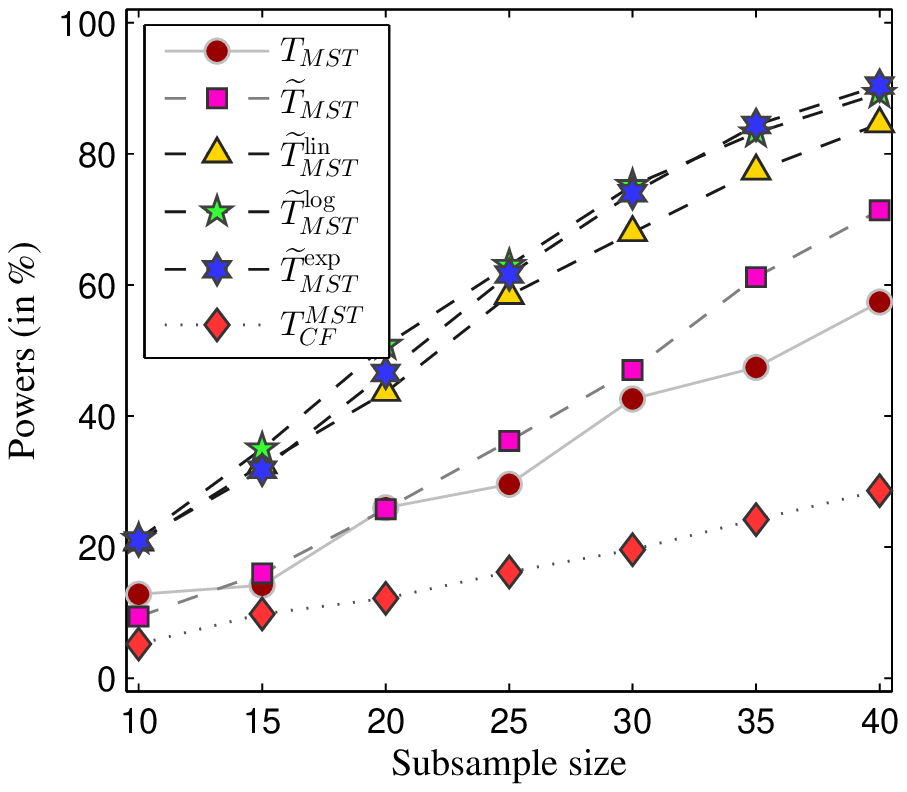}
\end{subfigure}
\caption{Powers of different tests in {\sc Lightning-2} data.}
\end{figure}

{\sc Gun-Point} data set comes from the video surveillance domain. This data set contains $100$ observations from each of two classes: {\sl Gun-Draw} and {\sl Point}. For {\sl Gun-Draw}, an actor draws a gun from a hip-mounted holster, points it at a target for approximately one second, and then returns the gun to the holster. For {\sl Point}, the actor does the same move, but instead of the gun, points the index finger to the target for approximately one second, and then returns to the initial position. For each class, an observation consists of $150$ measurements corresponding to the $X$ co-ordinate of the centroid of the actor's right hand during one movement. In this data set, modified NN tests based on different versions of MADD performed better than the NN test based on the Euclidean distance and the CF-NN test. Among these modified NN tests, the one based on $\rho_0$ had a slight edge. MST-run tests based on MADD and the Euclidean distance had almost similar powers. The overall performance of the CF-MST test was slightly inferior compared to other tests based on MST.

{\sc Lightning-2} data set contains observations from two classes: {\sl Cloud-to-Ground} lightning and {\sl Intra-Cloud} lightning. Each observation corresponds to transient electromagnetic events detected by FORTE satellite. Every input went through a Fourier transform to get a spectrogram, which was then collapsed in frequency to produce a power density time series. These time series were smoothed to produce $637$-dimensional observations. The data set consists of $48$ and $73$ observations from the two classes. Figure 12 shows the superiority of MADD based tests in this example. Modified NN and MST-run tests based $\rho_1$, $\rho_2$ and $\rho_3$ had much higher powers than their counterparts based on the Euclidean distance. Among them, the ones based on $\rho_2$ and $\rho_3$ outperformed others. Tests based on $\rho_0$ and those based on the Euclidean distance had almost similar performance. CF tests did not have satisfactory performance in this example. Powers of the CF-NN test (respectively, the CF-MST test) were much lower than all other NN tests (respectively, MST-run tests) considered here.

\section{Concluding Remarks}
\label{sec:remarks}

In this article, we have used {\MADD}, a new class of dissimilarity indices, to modify NN and MST-run tests. But the general recipe based on {\MADD} can also be used to improve the high-dimensional performance of many other two-sample tests. For instance, we can modify the tests based on averages of pairwise distances \citep[see, e.g.,][]{BF04,BF10,BG14,T17}, the SHP-run test \citep{BMG14} and the NBP test \citep{R05}. CF tests \citep{CF17} can be modified as well. High-dimensional consistency of the resulting tests can be proved using arguments similar to those used in this article. Using similar ideas, several multi-sample tests can also be modified to achieve better performance in high dimensions. For the NN test and its modified versions, throughout this article, we have reported all the numerical results for $k=3$ only. However, our findings remained almost the same for other values of $k$ as well. This is expected in view of the theoretical results stated in this article.

For the construction of the general version of {\MADD} (see Equation (2)),  we have used transformation on each of the measurement variables. Instead, one can partition the measurement vector $\xvec$ into $K$ non-overlapping blocks ${\tilde \xvec}_1,\ldots,{\tilde \xvec}_{K}$ of sizes $d_1,\ldots,d_K$ ($\sum_{i=1}^{K} d_i=d$), respectively, and define {\MADD} using blocked distance functions of the form $\varphi_{h,\psi}^{B}(\xvec,\yvec)$ = $h\{K^{-1} \sum_{q=1}^{K} \psi(\|{\tilde \xvec}_{q}-{\tilde \yvec}_{q}\|)\}$. As long as the block sizes are uniformly bounded, and the two distributions have different block distributions, consistency of the resulting tests based on MADD can be proved under conditions similar to Assumptions~\ref{assumption1}--\ref{assumption6}. This type of blocking can reveal more minute differences between two distributions. For instance, using blocks of size $2$, one can distinguish between two distributions having the same univariate marginals but different correlation structures. In that case, ideally, one would like to put highly correlated variables in the same block. In general, we would like to find blocks which are nearly independent, but the variables inside a block have significant dependence among themselves. But, at this moment, it is not yet clear how to develop an algorithm for finding such optimal blocks from the data. This can be considered as an interesting problem for future research.

\vspace{-0.1in}
\appendix
\section*{Proofs and mathematical details}

Throughout this section, we use $\Pr^\ast$ to denote conditional probability given ${\cal Z}_N$. So, we use $\Pr^\ast(A)$ to denote $\Pr(A \mid {\cal Z}_N)$ for an event $A$. For NN, MST-run, SHP-run and NBP tests based on $\varphi_{h,\psi}$, the tests statistics are denoted as $T_{NN}^{h,\psi}$, $T_{MST}^{h,\psi}$, $T_{SHP}^{h,\psi}$ and $T_{NBP}^{h,\psi}$, respectively.

\vspace{0.05in}
\noindent
{\bf Proof of Lemma~\ref{lemma:conv_Euclid}:}
Under Assumptions~\ref{assumption1} and \ref{assumption2}, for $\Wvec=\Xvec_1-\Xvec_2$, $\Yvec_1-\Yvec_2$ or $\Xvec_1-\Yvec_1$, $d^{-1} \sum_{q=1}^d \{{W^{(q)}}^2 - \E({W^{(q)}}^2)\}$ converges in probability to $0$. Also, $\sum_{q=1}^d \E({W^{(q)}}^2) = 2 tr(\sigmat_F)$, $2 tr(\sigmat_G)$ or $tr(\sigmat_F) + tr(\sigmat_G) + \|\muvec_F-\muvec_G\|^2$ for $\Wvec=\Xvec_1-\Xvec_2$, $\Yvec_1-\Yvec_2$ or $\Xvec_1-\Yvec_1$, respectively. Thus, under Assumption~\ref{assumption3}, as $d$ increases, $d^{-1} \|\Wvec\|^2$ converges in probability to $2 \sigma_F^2$, $2 \sigma_G^2$ or $\sigma_F^2 + \sigma_G^2 + \nu^2$ for $\Wvec = \Xvec_1-\Xvec_2$, $\Yvec_1-\Yvec_2$ or $\Xvec_1-\Yvec_1$, respectively. The proof now follows using the continuous mapping theorem. \qed

\vspace{0.05in}
\noindent
{\bf Proof of Lemma~\ref{lemma:separation}:} Let $\Xvec_1,\Xvec_2 \sim F$ and $\Yvec_1,\Yvec_2 \sim G$ be independent random vectors. Since $\psi^{\prime}(t)/t$ is a non-constant, monotone function,  for $q=1,\ldots,d$, we have $2 \E\psi(|X_1^{(q)}-Y_1^{(q)}|) - \E\psi(|X_1^{(q)}-X_2^{(q)}|) - \E\psi(|Y_1^{(q)}-Y_2^{(q)}|) \ge 0$, where the equality holds if and only if the $q$-th univariate marginals of $F$ and $G$ are the same \citep[see, e.g.,][]{BMG15}. As a result, one gets
\begin{equation}\label{eqn:lemma2_1}
\frac{2}{d} \sum_{q=1}^d \E\psi(|X_1^{(q)}-Y_1^{(q)}|) - \frac{1}{d} \sum_{q=1}^d \E\psi(|X_1^{(q)}-X_2^{(q)}|) - \frac{1}{d} \sum_{q=1}^d \E\psi(|Y_1^{(q)}-Y_2^{(q)}|) \ge 0,
\end{equation}
where the equality holds if and only if all univariate marginals of $F$ and $G$ are the same. Now, since $h$ is a concave and strictly increasing function, for any three real numbers $a,b$ and $c$ satisfying $2c-a-b \ge 0$, we have
\begin{equation}\label{eqn:lemma2_2}
h(c) \ge h\left(\frac{a+b}{2}\right) \ge \frac{1}{2} h(a) + \frac{1}{2} h(b) \Rightarrow 2h(c) - h(a) - h(b) \ge 0.
\end{equation}
The proof of the Lemma follows from Equations~\eqref{eqn:lemma2_1}~and~\eqref{eqn:lemma2_2}. \qed

\vspace{0.05in}
\noindent
{\bf Proof of Theorem~\ref{thm:SHP+NBP_consistent}:} $(a)$ Since $h$ is uniformly continuous, under Assumption~\ref{assumption4}, we have
\begin{equation}\label{eqn:thm1_1}
\varphi_{h,\psi}(\Xvec_i,\Xvec_j) - a_d \overset{\Pr}{\rightarrow} 0, \varphi_{h,\psi}(\Yvec_i,\Yvec_j) - b_d \overset{\Pr}{\rightarrow} 0 ~\text{and}~ \varphi_{h,\psi}(\Xvec_i,\Yvec_j) - c_d \overset{\Pr}{\rightarrow} 0 ~\text{as}~ d \to \infty,
\end{equation}
where $a_d=\varphi_{h,\psi}^{\ast}(F,F)$, $b_d=\varphi_{h,\psi}^{\ast}(G,G)$ and $c_d=\varphi_{h,\psi}^{\ast}(F,G)$ (see the discussion before Lemma~\ref{lemma:separation}). Since $\liminf_{d \to \infty} (2c_d - a_d - b_d) = \liminf_{d \to \infty} e_{h,\psi}(F,G) > 0$, following the proof of Theorem 1 in \cite{BMG14}, it is easy to show that $\Pr\big(T_{SHP}^{h,\psi} \le 3\big) \rightarrow 1$ as $d \to \infty$, where $T_{SHP}^{h,\psi}$ is the test statistic for the SHP-run test based on $\varphi_{h,\psi}$. Under ${\cal H}_0$, $T_{SHP}^{h,\psi}$ is distribution-free, and $\Pr_{{\cal H}_0}\big(T_{SHP}^{h,\psi} \le 3\big)= m!\,n!/(m+n-1)! = N/\binom{N}{m} < \alpha$ implies that the cutoff is larger than 3. This completes the proof.

\noindent
$(b)$ For the NBP test, first assume that $N$ is even. In that case, either $(i)$ both $m$ and $n$ are even or $(ii)$ both $m$ and $n$ are odd. In case $(i)$, the test statistic $T_{NBP}^{h,\psi}$ can take only even values, say $2k$. So, there are $2k$ pairs of the $\Xvec\Yvec$-type, $(m-2k)/2$ pairs of the $\Xvec\Xvec$-type and $(n-2k)/2$ pairs of the $\Yvec\Yvec$-type. If $\Delta^{m,n}_{2k,d} = \sum_{i=1}^{N/2} \varphi_{h,\psi}(\Zvec_{i1},\Zvec_{i2})$ denotes the corresponding total weight, then $\Delta^{m,n}_{2k,d} - C_{2k,d} \overset{\Pr}{\rightarrow} 0$ as $d \to \infty$, where $C_{2k,d} = (m-2k)a_d/2 + (n-2k)b_d/2 + 2kc_d = (2c_d-a_d-b_d)k + m a_d/2 + n b_d/2$ (see Equation~\eqref{eqn:thm1_1}). Since $\liminf_{d \to \infty} (2c_d-a_d-b_d)>0$, for all large $d$, this value is minimized for $k=0$. So, $T_{NBP}^{h,\psi} \overset{\Pr}{\rightarrow} 0$ as $d \to \infty$. In case $(ii)$, $T_{NBP}^{h,\psi}$ can take only odd values, say $2k-1$. Here also, one can check that $\Delta^{m,n}_{2k-1,d} - C_{2k-1,d} \overset{\Pr}{\rightarrow} 0$ as $d \to \infty$, where $C_{2k-1,d} = (k-1)(2c_d-a_d-b_d) + m a_d/2 + n b_d/2$. For all large $d$, $C_{2k-1,d}$ is minimized for $k=1$. So, $T_{NBP}^{h,\psi} \overset{\Pr}{\rightarrow} 1$ as $d \to \infty$. Under ${\cal H}_0$, $T_{NBP}^{h,\psi}$ is distribution-free, and following \cite{R05}, one can show that under the conditions on $m$ and $n$, both $\Pr_{{\cal H}_0}\big(T_{NBP}^{h,\psi} \le 0\big)$ in case $(i)$ and $\Pr_{{\cal H}_0}\big(T_{NBP}^{h,\psi} \le 1\big)$ in case $(ii)$ are less than $\alpha$. This completes the proof.

Now consider the case when $N$ is odd. Without loss of generality, let $m$ be odd and $n$ be even. Since $N$ is odd, one observation remains unpaired, and it is removed from the data. There are two possibilities, $(i)$ $m$ is reduced to $m-1$ and $(ii)$ $n$ is reduced to $n-1$. If case $(i)$ happens, since both $m-1$ and $n$ are even, following our discussion in the previous paragraph, for all large $d$, the total weight is minimized for $k=0$ and $\Delta^{m-1,n}_{0,d} - A_{0,d} \overset{\Pr}{\rightarrow} 0$ as $d \to \infty$, where $A_{0,d}=(m-1)a_d/2 + n b_d/2$. Similarly, if case $(ii)$ happens, since both $m$ and $n-1$ are odd, for all large $d$, the total weight is minimized for $k=1$ and $\Delta^{m,n-1}_{1,d}-A_{1,d} \overset{\Pr}{\rightarrow} 0$ as $d \to \infty$, where $A_{1,d} = A_{0,d} + c_d$. Note that $2c_d \ge e_{h,\psi}(F,G)$, and hence $\liminf_{d \to \infty} c_d > 0$. So, $A_{0,d}$ is strictly smaller than $A_{1,d}$ for all large $d$. So, case $(i)$ happens with probability tending to unity, and hence $T_{NBP}^{h,\psi} \overset{\Pr}{\rightarrow} 0$ as $d \to \infty$. Now, under the condition on $m$ and $n$, $\Pr_{{\cal H}_0} \big(T_{NBP}^{h,\psi} \le 0\big) = c(m,n) < \alpha$. This completes the proof. \qed

\vspace{0.05in}
\noindent
{\bf Proof of Theorem~\ref{thm:NN+MST_consistent}:} $(a)$ Note that the NN test is based on $T_{NN}^{h,\psi} = \frac{1}{Nk} \big[\sum_{i=1}^{m} \sum_{t=1}^k \bI_{t}^{h,\psi}(\Xvec_i)+ \sum_{j=1}^{n} \sum_{t=1}^k \bI_{t}^{h,\psi}(\Yvec_j)\big]$, where $\bI_{t}^{h,\psi}(\Zvec)$ is an indicator variable that takes the value $1$ if $\Zvec$ and its $t$-th nearest-neighbor in terms of $\varphi_{h,\psi}$ are from the same distribution. Recall that since $h$ is uniformly continuous, under Assumption~\ref{assumption4}, $\varphi_{h,\psi}(\Xvec_i,\Xvec_j) - a_d \overset{\Pr}{\rightarrow} 0$, $\varphi_{h,\psi}(\Yvec_i,\Yvec_j) - b_d \overset{\Pr}{\rightarrow} 0$ and $\varphi_{h,\psi}(\Xvec_i,\Yvec_j) - c_d \overset{\Pr}{\rightarrow} 0$ as $d \to \infty$, where $a_d = \varphi_{h,\psi}^{\ast}(F,F)$, $b_d = \varphi_{h,\psi}^{\ast}(G,G)$ and $c_d = \varphi_{h,\psi}^{\ast}(F,G)$ (see Equation~\eqref{eqn:thm1_1}). Since $\liminf_{d \to \infty} (c_d - a_d) > 0$ and $\liminf_{d \to \infty} (c_d - b_d) >0$, it follows that
\vspace{-0.1in}
\begin{align}
\label{eqn:thm4_1}
& \Pr\Big\{\max_{i \ne j} \varphi_{h,\psi}(\Xvec_i,\Xvec_j) < \min_{i,j} \varphi_{h,\psi}(\Xvec_i,\Yvec_j)\Big\} \to 1 ~\text{ and } \nonumber\\
& \Pr\Big\{\max_{i \ne j} \varphi_{h,\psi}(\Yvec_i,\Yvec_j) < \min_{i,j} \varphi_{h,\psi}(\Xvec_i,\Yvec_j)\Big\} \to 1 ~\text{ as } d \to \infty.
\end{align}
\vspace{-0.1in}

\noindent
So, for every $t \le k$, $\bI_{t}^{h,\psi}(\Xvec_i) \overset{\Pr}{\rightarrow} 1$ for $i=1,\ldots,m$ and $\bI_{t}^{h,\psi}(\Yvec_j) \overset{\Pr}{\rightarrow} 1$ for $j=1,\ldots,n$ as $d \to \infty$. Thus, $T_{NN}^{h,\psi}$ converges in probability to its maximum value $1$. Now, to prove the consistency of the test based on $T_{NN}^{h,\psi}$, we shall show that $\Pr^\ast\big(T_{NN}^{h,\psi}=1\big)<\alpha$ for almost every ${\cal Z}_N$.

Call $S \subseteq {\cal Z}_N$ to be a neighbor-complete set if for any $\zvec \in S$, all of its $k$ nearest-neighbors based on $\varphi_{h,\psi}$ also belong to $S$, and no proper subset of $S$ has this property. Clearly, $k+1 \le |S| \le N$, where $|S|$ denotes the cardinality of $S$. Let ${\cal Z}_N$ be partitioned into $r$ such neighbor-complete sets, i.e., ${\cal Z}_N= S_1 \cup \ldots \cup S_r$, 
where $r \le \lfloor N/(k+1) \rfloor$. Note that $T_{NN}^{h,\psi}=1$ if and only if, for each $i=1,\ldots,r$, all observations in $S_i$ have the same label. If $r_1 (<r)$ of these $S_i$'s are labelled $F$ and the rest are labelled $G$, then the sum of cardinalities of these $r_1$ sets should be $m$. Let $c_0(m,n)$ be the number of ways in which this can be done. Clearly, $\Pr^\ast(T_{NN}^{h,\psi}=1)= c_0(m,n)/ \binom{N}{m}$. So, it is enough to show that $c_0(m,n) \le \binom{N_0}{m_0}$.

First observe that we cannot have $T_{NN}^{h,\psi}=1$ if $N<2(k+1)$. If $N=2(k+1)$, it is possible only when $m=n=k+1$ and ${\cal Z}_N=S_1\cup S_2$, with $|S_1|=|S_2|=k+1$. So, in that case, all observations either in $S_1$ or in $S_2$ must be labelled as $F$. This leads to $c_0(m,n)=2$, and the result holds for $N=2(k+1)$.

Now, we shall prove the result using the method of mathematical induction on $N$. First assume that the result holds for all $N$ with $2(k+1) \le N \le M$. Without loss of generality, let us also assume that $m \le n$. For $N=M+1$, first note that observations in $S_1$ may or may not be labelled as $F$. Therefore, if $|S_1|=k_1$, we have $c_0(m,n)=c_0(m-k_1,n)+c_0(m,n-k_1)$. So, using the result for $N-k_1$, we get
\vspace{-0.05in}
$$c_0(m,n) \le \Bigg(\begin{array}{c}{\big\lceil \frac{N-k_1}{k+1}\big\rceil}\\ {\big\lceil \frac{m-k_1}{k+1}\big\rceil} \end{array}\Bigg) + \Bigg(\begin{array}{c}{\Big\lceil \frac{N-k_1}{k+1}\Big\rceil}\\ {\Big\lceil \frac{m}{k+1}\Big\rceil} \end{array} \Bigg).
\vspace{-0.05in}
$$
Here $\lceil (N-k_1)/(k+1) \rceil \le \lceil N/(k+1) \rceil - \lfloor k_1/(k+1) \rfloor \le N_0-1$ and $\lceil m/(k+1) \rceil=m_0$. So,
\vspace{-0.05in}
$$c_0(m,n) \le \Bigg(\begin{array}{c}{N_0-1}\\ {\big\lceil \frac{m-k_1}{k+1}\big\rceil} \end{array}
\Bigg) + \Bigg(\begin{array}{c}{N_0-1}\\ {m_0} \end{array} \Bigg).
\vspace{-0.05in}
$$
Also, observe that $\lceil (m-k_1)/(k+1) \rceil \le \lceil m/(k+1) \rceil - 1 = m_0 -1$ and $m_0-1 \le (N_0-1)/2$. Thus, $c_0(m,n) \le \binom{N_0-1}{m_0-1} + \binom{N_0-1}{m_0} = \binom{N_0}{m_0}$.

\noindent
$(b)$ {Note that the MST-run test based on $\varphi_{h,\psi}$ uses the test statistic $T_{MST}^{h,\psi} = 1 + \sum_{i=1}^{n-1} \lambda_i$, where $\lambda_{i}$ is an indicator variable that takes the value $1$ if the $i$-th edge of the MST on the complete graph with edge weights defined using $\varphi_{h,\psi}$ connects two observations from different distributions.} From Equation~\eqref{eqn:thm4_1}, it follows that for sufficiently large $d$, the MST on the vertex set ${\cal Z}_n$ has a sub-tree ${\cal T}_1$ on vertices corresponding to $m$ observations from $F$ and another sub-tree ${\cal T}_2$ on vertices corresponding to $n$ observations from $G$. These two sub-trees are connected by an edge of the $\Xvec\Yvec$-type \citep[see][]{BMG14}. As a result, $T_{MST}^{h,\psi}$ converges in probability to its minimum value $2$. From the proof of Theorem 2 in \cite{BMG14}, it follows that for sufficiently large $d$, $\Pr^\ast\big(T_{MST}^{h,\psi} \le 2\big) \le \max\{\lfloor N/m \rfloor, \lfloor N/n \rfloor\}/\binom{N}{m} < \alpha$ for almost every ${\cal Z}_N$. This proves the result. \qed

\vspace{0.05in}
\noindent
{\bf Proof of \autoref{thm:NN+MST_inconsistent}:} $(a)$ Recall that since $h$ is uniformly continuous, under Assumption~\ref{assumption4}, we have $\varphi_{h,\psi}(\Xvec_i,\Xvec_j) - a_d \overset{\Pr}{\rightarrow} 0$, $\varphi_{h,\psi}(\Yvec_i,\Yvec_j) - b_d \overset{\Pr}{\rightarrow} 0$ and $\varphi_{h,\psi}(\Xvec_i,\Yvec_j) - c_d \overset{\Pr}{\rightarrow} 0$, where $a_d = \varphi_{h,\psi}^{\ast}(F,F)$, $b_d = \varphi_{h,\psi}^{\ast}(G,G)$ and $c_d = \varphi_{h,\psi}^{\ast}(F,G)$ (see Equation~\eqref{eqn:thm1_1}). Since $2c_d - a_d - b_d \ge 0$ and $\limsup_{d \to \infty} (c_d - a_d) < 0$, it follows that $\liminf_{d \to \infty} (c_d - b_d) > 0$, and hence
\vspace{-0.05in}
\begin{align}
\label{eqn:thm5_1}
& \Pr\Big\{\max_{i \ne j} \varphi_{h,\psi}(\Yvec_i,\Yvec_j) < \min_{i,j} \varphi_{h,\psi}(\Xvec_i,\Yvec_j)\Big\} \to 1 ~\text{ and } \nonumber\\
& \Pr\Big\{\max_{i,j} \varphi_{h,\psi}(\Xvec_i,\Yvec_j) < \min_{i \ne j} \varphi_{h,\psi}(\Xvec_i,\Xvec_j)\Big\} \to 1 ~\text{ as } d \to \infty.
\end{align}
\vspace{-0.1in}

\noindent
As a result, for every $t \le k$, $\bI_{t}^{h,\psi}(\Xvec_i) \overset{\Pr}{\rightarrow} 0$ for $i=1,\ldots,m$ and $\bI_{t}^{h,\psi}(\Yvec_j) \overset{\Pr}{\rightarrow} 1$ for $j=1,\ldots,n$ as $d \to \infty$. Thus, $T_{NN}^{h,\psi} \overset{\Pr}{\rightarrow} n/N$ as $d \to \infty$. Now, from the proof of Theorem 3.2(b) in \cite{BG14}, it follows that when $(m-1)/n > (1+\alpha)/(1-\alpha)$, $\Pr^{\ast}\big(T_{NN}^{h,\psi} \ge n/N\big) < \alpha$ for almost every ${\cal Z}_N$. This proves part $(a)$ of the theorem.

\noindent
$(b)$ Equation~\eqref{eqn:thm5_1} implies that $T_{MST}^{h,\psi} \overset{\Pr}{\rightarrow} m+1$ as $d \to \infty$ \citep[see][]{BG14}. Under the condition $m/n > (1+\alpha)/(1-\alpha)$, from the proof of Theorem 2(ii) in \cite{BMG14}, it also follows that $\Pr^{\ast}\big(T_{MST}^{h,\psi} \le m+1\big) \ge (m-n)/N > \alpha$ for almost every ${\cal Z}_N$. Thus, the cutoff obtained using the permutation principle turns out to be strictly smaller than the observed value with probability converging to unity as the dimension increases. This completes the proof. \qed

\vspace{0.05in}
\noindent
{\bf Proof of Lemma~\ref{lemma:semi-metric}:} Symmetry and non-negativity of $\rho_{h,\psi}$ are obvious. So, we shall prove the triangle inequality for $\rho_{h,\psi}$. First observe that
\begin{align}
\bigl|\varphi_{h,\psi}(\zvec_1,\zvec_3)-\varphi_{h,\psi}(\zvec_2,\zvec_3)\bigr| & = \bigl|\varphi_{h,\psi}(\zvec_1,\zvec_2)-\varphi_{h,\psi}(\zvec_2,\zvec_3)-\varphi_{h,\psi}(\zvec_1,\zvec_2)+\varphi_{h,\psi}(\zvec_1,\zvec_3)\bigr| \nonumber \\
& \le \bigl|\varphi_{h,\psi}(\zvec_1,\zvec_2)-\varphi_{h,\psi}(\zvec_3,\zvec_2)\bigr|+\bigl|\varphi_{h,\psi}(\zvec_2,\zvec_1)
-\varphi_{h,\psi}(\zvec_3,\zvec_1)\bigr|. \nonumber
\end{align}
This proves the result for $N=3$. If $N \ge 4$, for any $\zvec_k$ with $k \ge 4$,
\begin{align}
\bigl|\varphi_{h,\psi}(\zvec_1,\zvec_k)-\varphi_{h,\psi}(\zvec_2,\zvec_k)\bigr| & = \bigl|\varphi_{h,\psi}(\zvec_1,\zvec_k)-\varphi_{h,\psi}(\zvec_3,\zvec_k)+\varphi_{h,\psi}(\zvec_3,\zvec_k)-\varphi_{h,\psi}(\zvec_2,\zvec_k)\bigr| \nonumber \\
& \le \bigl|\varphi_{h,\psi}(\zvec_1,\zvec_k)-\varphi_{h,\psi}(\zvec_3,\zvec_k)\bigr|+\bigl|\varphi_{h,\psi}
(\zvec_2,\zvec_k)-\varphi_{h,\psi}(\zvec_3,\zvec_k)\bigr|. \nonumber
\end{align}
Combining these above-mentioned inequalities, we get
\begin{align*}
\sum_{k \ne {1,2}} \bigl|\varphi_{h,\psi}(\zvec_1,\zvec_k)-\varphi_{h,\psi}(\zvec_2,\zvec_k)\bigr| &\le \sum_{k \ne {1,3}} \bigl|\varphi_{h,\psi}(\zvec_1,\zvec_k)-\varphi_{h,\psi}(\zvec_3,\zvec_k)\bigr|\\
&~~~~~~~~~~~~~~~~+\sum_{k \ne {2,3}} \bigl|\varphi_{h,\psi}(\zvec_2,\zvec_k)-\varphi_{h,\psi}(\zvec_3,\zvec_k)\bigr|.
\end{align*}
This implies $\rho_{h,\psi}(\zvec_1,\zvec_2) \le \rho_{h,\psi}(\zvec_1,\zvec_3) + \rho_{h,\psi}(\zvec_2,\zvec_3)$. \qed

\vspace{0.05in}
\noindent
{\bf Proof of Lemma~\ref{lemma:conv_MADD}:} Following Lemma~\ref{lemma:conv_Euclid}, under Assumptions~\ref{assumption1}--\ref{assumption3}, $d^{-1/2} \|\Xvec_1-\Xvec_2\|$,   $d^{-1/2} \|\Yvec_1-\Yvec_2\|$ and  $d^{-1/2} \|\Xvec_1-\Yvec_1\|$ converge in probability to $\sigma_F\sqrt{2}$, $\sigma_G\sqrt{2}$ and $\sqrt{\sigma_F^2+\sigma_G^2+\nu^2}$, respectively, as $d$ tends to infinity. Since $m$ and $n$ are finite, $d^{-1/2} \rho_0(\Xvec_1,\Xvec_2)$ and $d^{-1/2} \rho_0(\Yvec_1,\Yvec_2)$ have probability convergence to $0$, while  $d^{-1/2} \rho_0(\Xvec_1,\Yvec_1)$ converges in probability to $\widetilde{\rho}_0(F,G) = (N-2)^{-1} \big\{(m-1) \big|\sigma_F\sqrt{2} - \sqrt{\sigma_F^2 + \sigma_G^2 + \nu^2}\big| + (n-1) \big|\sqrt{\sigma_F^2 + \sigma_G^2 + \nu^2} - \sigma_G\sqrt{2}\big|\big\}$. Clearly, $\widetilde{\rho}_0(F,G) \ge 0$, where equality holds if and only if $\sigma_F^2 = \sigma_G^2 + \nu^2$ and $\sigma_G^2 = \sigma_F^2 + \nu^2$, i.e., $\nu^2 = 0$ and $\sigma_F^2 = \sigma_G^2$. \qed

\vspace{0.05in}
\noindent
{\bf Proof of Theorem~\ref{thm:rho0_consistency}:} Following Lemma~\ref{lemma:conv_MADD}, under Assumptions~\ref{assumption1}--\ref{assumption3}, we have
\begin{equation*}
\Pr\big[\min_{i,j}\rho_0(\Xvec_i,\Yvec_j) > \max\big\{\max_{i \ne j}\rho_0(\Xvec_{i},\Xvec_{j}), \max_{i \ne j} \rho_0(\Yvec_{i},\Yvec_{j})\big\}\big] \to 1 \text{ as } d \to \infty.
\end{equation*}
The rest of the proof is similar to the proof of Theorem~\ref{thm:NN+MST_consistent}. \qed

\noindent
{\bf Proof of Lemma~\ref{lemma:MADD_general}:}
Since $h$ is strictly increasing, $\rho_{h,\psi}^\ast(F,G)=0$ implies $\sum_{q=1}^d \E\psi(|X_1^{(q)}-X_2^{(q)}|) = \sum_{q=1}^d \E\psi(|Y_1^{(q)}-X_2^{(q)}|)$ and $\sum_{q=1}^d \E\psi(|X_1^{(q)}-Y_2^{(q)}|) = \sum_{q=1}^d \E\psi(|Y_1^{(q)}-Y_2^{(q)}|)$. So, $\sum_{q=1}^{d} e_{F,G}^{(q)}$ $=\sum_{q=1}^d \big\{2\E\psi(|X_1^{(q)}-Y_1^{(q)}|) - \E\psi(|X_1^{(q)}-X_2^{(q)}|) - \E\psi(|Y_1^{(q)}-Y_2^{(q)}|)\big\} = 0$. Since $\psi^{\prime}(t)/t$ is a non-constant, monotone function, for each $q=1,\ldots,d$, $e_{F,G}^{(q)}$ is non-negative and it takes the value $0$ if and only if the $q$-th marginal distributions of $F$ and $G$ are the same  \citep[see][]{BF10,BMG15}. Thus, $\rho_{h,\psi}^{\ast}(F,G) = 0$ implies that $F$ and $G$ have the same univariate marginal distributions. On the other hand, when $F$ and $G$ have the same univariate marginal distributions, it follows trivially that $\varphi_{h,\psi}^\ast(F,F) = \varphi_{h,\psi}^\ast(G,G) = \varphi_{h,\psi}^\ast(F,G)$, and hence $\rho_{h,\psi}^\ast(F,G)=0$. \qed

\vspace{0.05in}
\noindent
{\bf Proof of Theorem~\ref{thm:Power_MADDgeneral}:}
The proof is similar to the proofs of Theorems~\ref{thm:NN+MST_consistent}~and~\ref{thm:rho0_consistency} with the use of Assumption~\ref{assumption5}. Hence we skip the details of the proof. \qed

\vspace{0.05in}
\noindent
{\bf Proof of Theorem~\ref{thm:MADD_weaker}:}
Consider independent random vectors $\Xvec_1,\Xvec_2 \sim F$, $\Yvec_1,\Yvec_2 \sim G$ and $\Zvec \sim H$, where $H=F$ or $G$. Define $S_{d}= d^{-1}\sum_{q=1}^d \psi(|X_1^{(q)}-Y_1^{(q)}|)$. Since $\big(S_{d}-\E(S_{d})\big)/\sqrt{\var(S_{d})}={\bf O}_{P}(1)$, we get $S_{d}-\E(S_{d})$ $={\bf O}_{P}(\vartheta(d)/d)$. As $h$ is Lipschitz continuous, 
\begin{equation*}
\big|\varphi_{h,\psi}(\Xvec_1,\Yvec_1) - \varphi_{h,\psi}^\ast(F,G)\big| = |h(S_d) - h\{\E(S_d)\}| \le C_0 |S_d - \E(S_d)| = {\bf O}_{P}(\vartheta(d)/d).
\end{equation*}
Similarly, $\big|\varphi_{h,\psi}(\Xvec_1,\Zvec) - \varphi_{h,\psi}^\ast(F,H)\big|$ and $\big|\varphi_{h,\psi}(\Yvec_1,\Zvec) - \varphi_{h,\psi}^\ast(G,H)\big|$ are also of the order ${\bf O}_{P}(\vartheta(d)/d)$. So, $\big|\varphi_{h,\psi}(\Xvec_1,\Zvec) - \varphi_{h,\psi}(\Yvec_1,\Zvec)\big| = \big|\varphi_{h,\psi}^{\ast}(F,H) - \varphi_{h,\psi}^{\ast}(G,H)\big| + {\bf O}_{P}(\vartheta(d)/d)$. Since $m$ and $n$ are finite, this implies $\rho_{h,\psi}(\Xvec_1,\Yvec_1)$ = $\rho_{h,\psi}^{\ast}(F,G)$ + ${\bf O}_{P}(\vartheta(d)/d)$. Similarly, we get $\rho_{h,\psi}(\Xvec_1,\Xvec_2) = {\bf O}_{P}(\vartheta(d)/d)$ and $\rho_{h,\psi}(\Yvec_1,\Yvec_2) = {\bf O}_{P}(\vartheta(d)/d)$. Under Assumption~\ref{assumption6}, $\rho_{h,\psi}^{\ast}(F,G)$ has asymptotic order higher than that of $\vartheta(d)/d$. So, $$\Pr\big[\rho_{h,\psi}(\Xvec_1,\Yvec_1) > \max\{\rho_{h,\psi}(\Xvec_1,\Xvec_2), \rho_{h,\psi}(\Yvec_1,\Yvec_2)\}\big] \rightarrow 1 ~\text{as}~ d \rightarrow \infty.$$
This proves the first part of the theorem. The consistency part now follows using arguments similar to those used in the proofs of Theorems~\ref{thm:NN+MST_consistent},~\ref{thm:rho0_consistency}~and~\ref{thm:Power_MADDgeneral}. \qed

\vspace{0.05in}
\noindent
{\bf Proof of Theorem~\ref{thm:Euclid_weaker}:}
For $\rho_0$, we use $h(t)=\sqrt{t}$ and $\psi(t) = t^2$. So, for $\Xvec_1 \sim F$ and $\Yvec_1 \sim G$, taking $S_d = d^{-1}\sum_{q=1}^d (X_1^{(q)}-Y_1^{(q)})^2$, we get 
\begin{equation*}
\varphi_{h,\psi}(\Xvec_1,\Yvec_1) - \varphi_{h,\psi}^{\ast}(F,G) = \sqrt{S_d} - \sqrt{\E(S_d)} = \frac{S_d-\E(S_d)}{\sqrt{S_d} + \sqrt{\E(S_d)}}.
\end{equation*}
Here $\E(S_d) = d^{-1}\big\{\|\muvec_F-\muvec_G\|^2 + tr(\sigmat_F+\sigmat_G)\big\} \ge d^{-1}tr(\sigmat_F)$. So, $\sqrt{d\E(S_d)/\vartheta(d)}$ remains bounded away from $0$, and hence $\sqrt{\vartheta(d)}/\big(\sqrt{d S_d}+\sqrt{d\,\E(S_d)}\big)$ remains bounded as $d$ goes to infinity. Now, $(S_d-\E(S_d))/\sqrt{\var(S_d)}={\bf O}_{P}(1)$ implies $S_d-\E(S_d)={\bf O}_{P}(\vartheta(d)/d)$. Again, $1/\big(\sqrt{S_d} + \sqrt{\E(S_d)}\big) = {\bf O}_{P}(\sqrt{d/\vartheta(d)})$. So, $\varphi_{h,\psi}(\Xvec_1,\Yvec_1) = \varphi_{h,\psi}^{\ast}(F,G) + {\bf O}_{P}(\sqrt{\vartheta(d)/d})$. Thus, as in the proof of Theorem~\ref{thm:MADD_weaker}, we get $\rho_0(\Xvec_1,\Yvec_1) = \rho_0^\ast(F,G) + {\bf O}_{P}(\sqrt{\vartheta(d)/d})$, where $\rho_0^\ast(F,G)$ is $\rho_{h,\psi}^\ast(F,G)$ with $h(t)=\sqrt{t}$ and $\psi(t)=t^2$. Similarly, we have $\rho_0(\Xvec_1,\Xvec_2) = {\bf O}_{P}(\sqrt{\vartheta(d)/d})$ and $\rho_0(\Yvec_1,\Yvec_2) = {\bf O}_{P}(\sqrt{\vartheta(d)/d})$. Now, it is easy to check that when $\|\muvec_F-\muvec_G\|^2/\vartheta(d)$ or $|tr(\sigmat_F)-tr(\sigmat_G)|/\vartheta(d)$ diverge to infinity, $\rho_0^\ast(F,G)$ is of higher order than $\sqrt{\vartheta(d)/d}$. The rest of the proof is similar to the proof of Theorem~\ref{thm:MADD_weaker}. \qed

\bibliographystyle{rss}
\bibliography{Bibliography}

\end{document}